\documentclass[hyper,a4paper,12pt,oneside]{JHEP3}

\title{Hamiltonian BRST deformation of a class of
$n$-dimensional BF-type theories}
\author{Constantin Bizdadea\\Faculty of Physics, University of Craiova\\
13 A. I. Cuza Str., Craiova RO-1100, Romania\\
E-mail: \email{bizdadea@central.ucv.ro}}
\author{Costin C\u{a}t\u{a}lin Ciob\^\i rc\u{a}\\Faculty of Physics,
University of Craiova\\
13 A. I. Cuza Str., Craiova RO-1100, Romania\\
E-mail: \email{ciobarca@central.ucv.ro}}
\author{Eugen Mih\u{a}i\c{t}\u{a} Cioroianu\\Faculty of Physics,
University of Craiova\\
13 A. I. Cuza Str., Craiova RO-1100, Romania\\
E-mail: \email{manache@central.ucv.ro}}
\author{Solange Odile Saliu\\Faculty of Physics, University of Craiova\\
13 A. I. Cuza Str., Craiova RO-1100, Romania\\
E-mail: \email{osaliu@central.ucv.ro}}
\author{Silviu Constantin S\u{a}raru\\Faculty of Physics, University of Craiova\\
13 A. I. Cuza Str., Craiova RO-1100, Romania\\
E-mail: \email{scsararu@central.ucv.ro}}

\abstract{Consistent Hamiltonian interactions that can be added to
an abelian free BF-type class of theories in any $n\geq 4$
spacetime dimensions are constructed in the framework of the
Hamiltonian BRST deformation based on cohomological techniques.
The resulting model is an interacting field theory in higher
dimensions with an open algebra of on-shell reducible first-class
constraints. We argue that the Hamiltonian couplings are related
to a natural structure of Poisson manifold on the target space.}


\preprint{}

\begin{document}

\section{Introduction}

The Hamiltonian BRST symmetry \cite{2,8} has the advantage of a
proper implementation in quantum mechanics \cite{2} (Chapter 14),
and also of an appropriate correlation with the canonical
quantization methods \cite {9}. The understanding of this symmetry
on cohomological grounds made possible a unitary approach to many
problems in gauge field theory, such as the Hamiltonian analysis
of anomalies \cite{10}, the precise relation between local
Lagrangian and Hamiltonian BRST cohomologies \cite{11}, and,
recently, the problem of obtaining consistent Hamiltonian
interactions in gauge theories by means of the deformation theory
\cite{12,mpla,ijmpa}.

In this paper we investigate the consistent Hamiltonian BRST deformations in
any spacetime dimension $n\geq 4$ of a free abelian topological field theory%
\footnote{%
By `topological field theory' we mean a gauge theory whose moduli
space is finite dimensional, but nonzero, in contrast to
`topological invariant theories', whose moduli space is zero
dimensional.} of BF-type \cite{13} involving a set of scalar
fields, two collections of one-forms and a system of two-forms.
Actually, this work generalizes our previous results in two and
four spacetime dimensions \cite{mpla,ijmpa}. We show that the
resulting interactions are accurately described by a topological
field theory with an open algebra of first-class constraints, that
can be interpreted in terms of a Poisson structure present in
various models of two-dimensional gravity
\cite{grav1,grav2,grav3,grav4}. The analysis of Poisson Sigma
Models, including their relationship to two-dimensional gravity
and the study of classical solutions, can be found in
\cite{psm1,stroblspec,psm2,psm3,psm4,psmn} (see also
\cite{ikeda12}). In view of this interpretation, we believe that
the open problem of how to approach the Hamiltonian quantization
of gravity without string theory might benefit from the
construction of consistently interacting Hamiltonian gauge
theories (in higher spacetime dimensions) with special classes of
open algebras.

The plan of the paper is the following. Section 2 briefly reviews the
problem of constructing consistent Hamiltonian interactions in the framework
of the BRST formalism, which reduces to solving two towers of equations that
describe the deformation of the BRST charge, respectively, of the
BRST-invariant Hamiltonian associated with a given ``free'' first-class
theory at various orders in the coupling constant. In Section 3 we determine
the Hamiltonian BRST symmetry ($s$) of the free theory under study in $n\geq
4$ spacetime dimensions, which splits as the sum between the Koszul-Tate
differential and the exterior derivative along the gauge orbits. This model
is abelian and $\left( n-2\right) $-stage reducible, the reducibility
relations holding off-shell. Next, we solve the main equations governing the
Hamiltonian deformation procedure on behalf of the BRST cohomology of the
free theory. In Section 4 we initially compute, using specific cohomological
techniques, the first-order deformation of the BRST charge, which lies in
the cohomological space of $s$ modulo the spatial part of the exterior
spacetime derivative ($\tilde{d}$) at ghost number one, $H^{1}\left( s|%
\tilde{d}\right) $. The first-order deformation of the BRST charge stops at
antighost number $\left( n-1\right) $ and contains two sorts of arbitrary
functions involving only the undifferentiated scalar fields. Its consistency
reveals one `two-tensor' (in the collection indices) depending on the scalar
fields, that must be antisymmetric and fulfill a certain identity. Under
these conditions, all the other deformations, of order two and higher, can
be taken to vanish, and thus the BRST charge of the interacting model that
is consistent to all orders in the deformation parameter is fully output.
Section 5 solves the problem of generating the deformed BRST-invariant
Hamiltonian, which can be taken nonzero only at the first order in the
coupling constant. With the help of these deformed quantities, in Section 6
we identify the interacting gauge theory, which displays an open algebra of
constraints. The deformed first-class constraints are of course reducible,
but the reducibility relations hold on-shell. In the meantime, we can
interpret the `two-tensor' mentioned in the above in terms of a target space
parametrized by the scalar fields and endowed with a natural structure of
Poisson manifold. Section 7 contains the main conclusions of the present
paper.

\section{Main equations of the Hamiltonian deformation procedure}

It has been shown in \cite{12} that the problem of constructing consistent
Hamiltonian interactions in theories with first-class constraints can be
equivalently reformulated as a deformation problem of the BRST charge $%
\Omega _{0}$ and of the BRST-invariant Hamiltonian $H_{0\mathrm{B}}$ of a
given ``free'' first-class theory. More precisely, if the interactions can
be consistently constructed, then the ``free'' BRST charge can be deformed
into
\begin{eqnarray}
&&\Omega _{0}\rightarrow \Omega =\Omega _{0}+g\int d^{n-1}x\,\omega
_{1}+g^{2}\int d^{n-1}x\,\omega _{2}+O\left( g^{3}\right)  \nonumber \\
&=&\Omega _{0}+g\Omega _{1}+g^{2}\Omega _{2}+O\left( g^{3}\right) ,
\label{d1}
\end{eqnarray}
where the BRST charge of the interacting theory, $\Omega $, must satisfy the
equation
\begin{equation}
\left[ \Omega ,\Omega \right] =0.  \label{d2}
\end{equation}
The symbol $[,]$ signifies the Poisson (Dirac) bracket. The last
formula projected on various powers in the deformation parameter
$g$ leads to the tower of equations
\begin{equation}
\left[ \Omega _{0},\Omega _{0}\right] =0,  \label{d3}
\end{equation}
\begin{equation}
2\left[ \Omega _{0},\Omega _{1}\right] =0,  \label{d4}
\end{equation}
\begin{equation}
2\left[ \Omega _{0},\Omega _{2}\right] +\left[ \Omega _{1},\Omega
_{1}\right] =0,  \label{d5}
\end{equation}
\[
\vdots
\]
In a similar manner, we deform the BRST-invariant Hamiltonian of the
``free'' theory
\begin{eqnarray}
&&H_{0\mathrm{B}}\rightarrow H_{\mathrm{B}}=H_{0\mathrm{B}}+g\int
d^{n-1}x\,h_{1}+g^{2}\int d^{n-1}x\,h_{2}+O\left( g^{3}\right)  \nonumber \\
&=&H_{0\mathrm{B}}+gH_{1}+g^{2}H_{2}+O\left( g^{3}\right) ,  \label{d6}
\end{eqnarray}
and require that it stands for the BRST-invariant Hamiltonian of the coupled
system
\begin{equation}
\left[ H_{\mathrm{B}},\Omega \right] =0.  \label{d7}
\end{equation}
The decomposition of the relation (\ref{d7}) according to the various orders
in the coupling constant reveals a new tower of equations
\begin{equation}
\left[ H_{0\mathrm{B}},\Omega _{0}\right] =0,  \label{d8}
\end{equation}
\begin{equation}
\left[ H_{1},\Omega _{0}\right] +\left[ H_{0\mathrm{B}},\Omega _{1}\right]
=0,  \label{d9}
\end{equation}
\begin{equation}
\left[ H_{2},\Omega _{0}\right] +\left[ H_{1},\Omega _{1}\right] +\left[ H_{0%
\mathrm{B}},\Omega _{2}\right] =0,  \label{d10}
\end{equation}
\[
\vdots
\]
While the equations (\ref{d3}) and (\ref{d8}) are satisfied since
$\Omega _{0}$ and $H_{0\mathrm{B}}$ are by hypothesis the BRST
charge, respectively, the BRST-invariant Hamiltonian of the
``free'' theory, the resolution of the remaining equations
((\ref{d4}--\ref{d5}), etc. and (\ref{d9}--\ref{d10}), etc.) by
means of cohomological techniques provides the Hamiltonian BRST
description of the deformed gauge theory.

\section{Free BRST symmetry}

Our starting point is a free topological field theory of BF-type in any $%
n\geq 4$ spacetime dimension, that involves two types of one-forms, a
collection of scalar fields, and a system of two-forms, described by the
Lagrangian action
\begin{equation}
S_{0}\left[ A_{\mu }^{a},H_{\mu }^{a},\varphi _{a},B_{a}^{\mu \nu }\right]
=\int d^{n}x\left( H_{\mu }^{a}\partial ^{\mu }\varphi _{a}+\frac{1}{2}%
B_{a}^{\mu \nu }\partial _{[\mu }A_{\nu ]}^{a}\right) ,  \label{f1}
\end{equation}
where here and in the sequel the notation $\left[ \mu \nu \right]
$ signifies antisymmetry with respect to the indices between
brackets. We work with the Minkowskian metric $g_{\mu \nu}$ of
`mostly minus' signature: $(+,---\cdots )$.

The above action is invariant under the gauge transformations
\begin{equation}
\delta _{\epsilon }A_{\mu }^{a}=\partial _{\mu }\epsilon ^{a},\;\delta
_{\epsilon }H_{\mu }^{a}=\partial ^{\nu }\epsilon _{\mu \nu }^{a},\;\delta
_{\epsilon }\varphi _{a}=0,\;\delta _{\epsilon }B_{a}^{\mu \nu }=\partial
_{\rho }\epsilon _{a}^{\mu \nu \rho },  \label{f2}
\end{equation}
which are off-shell $\left( n-2\right) $-stage reducible, where the gauge
parameters $\epsilon ^{a}$, $\epsilon _{\mu \nu }^{a}$ and $\epsilon
_{a}^{\mu \nu \rho }$ are bosonic, the last two sets being completely
antisymmetric.

We denote by $\left( \pi _{a}^{\mu },\pi _{\mu \nu }^{a},p_{a}^{\mu
},p^{a}\right) $ the canonical momenta respectively conjugated to the fields
$\left( A_{\mu }^{a},B_{a}^{\mu \nu },H_{\mu }^{a},\varphi _{a}\right) $. By
performing the canonical analysis and by eliminating the second-class
constraints (the coordinates of the reduced phase-space are $z^{A}=\left(
\pi _{a}^{0},A_{\mu }^{a},B_{a}^{\mu \nu },p_{a}^{i},H_{\mu }^{a},\pi
_{ij}^{a},\varphi _{a}\right) $), we are left with a system subject only to
the first-class constraints
\begin{equation}
G_{a}^{(1)}\equiv \pi _{a}^{0}\approx 0,\;G_{a}^{(2)}\equiv -\partial
_{i}B_{a}^{0i}\approx 0,  \label{f3}
\end{equation}
\begin{equation}
G_{ij}^{(1)a}\equiv 2\pi _{ij}^{a}\approx 0,\;G_{ij}^{(2)a}\equiv -\partial
_{[i}A_{j]}^{a}\approx 0,  \label{f4}
\end{equation}
\begin{equation}
\gamma _{a}^{(1)i}\equiv -p_{a}^{i}\approx 0,\;\gamma _{a}^{(2)i}\equiv
\partial ^{i}\varphi _{a}\approx 0,  \label{f5}
\end{equation}
and displaying the first-class Hamiltonian
\begin{equation}
H_{0}=\int d^{n-1}x\left( -H_{i}^{a}\gamma _{a}^{(2)i}+\frac{1}{2}%
B_{a}^{ij}G_{ij}^{(2)a}+A_{0}^{a}G_{a}^{(2)}\right) ,  \label{f6}
\end{equation}
in terms of the non-vanishing fundamental Dirac brackets
\begin{equation}
\left[ \pi _{a}^{0}(t,\mathbf{x}),A_{0}^{b}(t,\mathbf{y})\right] =-\delta
_{a}^{b}\delta ^{n-1}\left( \mathbf{x-y}\right) ,  \label{f7}
\end{equation}
\begin{equation}
\left[ B_{a}^{0i}(t,\mathbf{x}),A_{j}^{b}(t,\mathbf{y})\right] =-\delta
_{j}^{i}\delta _{a}^{b}\delta ^{n-1}\left( \mathbf{x-y}\right) ,  \label{f8}
\end{equation}
\begin{equation}
\left[ H_{0}^{a}(t,\mathbf{x}),\varphi _{b}(t,\mathbf{y})\right] =-\delta
_{b}^{a}\delta ^{n-1}\left( \mathbf{x-y}\right) ,  \label{f9}
\end{equation}
\begin{equation}
\left[ \pi _{ij}^{a}(t,\mathbf{x}),B_{b}^{kl}(t,\mathbf{y})\right] =-\frac{1%
}{2}\delta _{i}^{[k}\delta _{j}^{l]}\delta _{b}^{a}\delta ^{n-1}\left(
\mathbf{x-y}\right) ,  \label{f10}
\end{equation}
\begin{equation}
\left[ p_{a}^{i}(t,\mathbf{x}),H_{j}^{b}(t,\mathbf{y})\right] =-\delta
_{j}^{i}\delta _{a}^{b}\delta ^{n-1}\left( \mathbf{x-y}\right) .  \label{f11}
\end{equation}
The above constraints are abelian, while the remaining gauge algebra
relations are expressed by
\begin{equation}
\left[ H_{0},G_{a}^{(1)}\right] =G_{a}^{(2)},\;\left[
H_{0},G_{a}^{(2)}\right] =0,  \label{f12}
\end{equation}
\begin{equation}
\left[ H_{0},G_{ij}^{(1)a}\right] =G_{ij}^{(2)a},\;\left[
H_{0},G_{ij}^{(2)a}\right] =0,  \label{f13}
\end{equation}
\begin{equation}
\left[ H_{0},\gamma _{a}^{(1)i}\right] =\gamma _{a}^{(2)i},\;\left[
H_{0},\gamma _{a}^{(2)i}\right] =0.  \label{f14}
\end{equation}
The constraint functions $G_{ij}^{(2)a}$ are off-shell $\left( n-3\right) $%
-stage reducible, with the reducibility functions (of order $(k-2)$) given
by
\begin{equation}
\left( Z_{i_{1}i_{2}\cdots i_{k}}^{a}\right) _{b}^{j_{1}\cdots j_{k-1}}=%
\frac{\left( -\right) ^{k-1}}{\left( k-1\right) !}\delta _{b}^{a}\partial
_{[i_{1}}\delta _{i_{2}}^{j_{1}}\cdots \delta
_{i_{k}]}^{j_{k-1}},\;k=3,\cdots ,n-1,  \label{f15}
\end{equation}
while the constraint functions $\gamma _{a}^{(2)i}$ are off-shell $\left(
n-2\right) $-stage reducible, the associated reducibility functions (of
order $(k-1)$) being
\begin{equation}
\left( Z_{a}^{i_{1}i_{2}\cdots i_{k}}\right) _{j_{1}\cdots j_{k-1}}^{b}=%
\frac{\left( -\right) ^{k-1}}{\left( k-1\right) !}\delta _{a}^{b}\partial
^{[i_{1}}\delta _{j_{1}}^{i_{2}}\cdots \delta
_{j_{k-1}}^{i_{k}]},\;k=2,\cdots ,n-1.  \label{f16}
\end{equation}

The Hamiltonian BRST formalism requires the introduction of the ghosts
\begin{equation}
\eta ^{a_{0}}=\left( \eta ^{(1)a},\eta ^{a},\eta _{a}^{(1)ij},\eta
_{a}^{ij},C_{i}^{(1)a},C_{i}^{a}\right) ,  \label{f17}
\end{equation}
\begin{equation}
\eta ^{a_{k}}=\left( C_{i_{1}\cdots i_{k+1}}^{a},\eta _{a}^{i_{1}\cdots
i_{k+2}}\right) ,\;k=1,\cdots ,n-3,  \label{f18}
\end{equation}
\begin{equation}
\eta ^{a_{n-2}}=\left( C_{i_{1}\cdots i_{n-1}}^{a}\right) ,  \label{f19}
\end{equation}
together with their conjugated antighosts
\begin{equation}
\mathcal{P}_{a_{0}}=\left( \mathcal{P}_{a}^{(1)},\mathcal{P}_{a},\mathcal{P}%
_{ij}^{(1)a},\mathcal{P}_{ij}^{a},P_{a}^{(1)i},P_{a}^{i}\right) ,
\label{f20}
\end{equation}
\begin{equation}
\mathcal{P}_{a_{k}}=\left( P_{a}^{i_{1}\cdots i_{k+1}},\mathcal{P}%
_{i_{1}\cdots i_{k+2}}^{a}\right) ,\;k=1,\cdots ,n-3,  \label{f21}
\end{equation}
\begin{equation}
\mathcal{P}_{a_{n-2}}=\left( P_{a}^{i_{1}i_{2}\cdots i_{n-1}}\right) .
\label{f22}
\end{equation}
The first set of ghosts respectively corresponds to the first-class
constraints (\ref{f3}--\ref{f5}), while the other two are due to the
reducibility of the first-class constraint functions. The fields $\eta
^{a_{0}}$ in (\ref{f17}) are fermionic and of ghost number one, the fields $%
\eta ^{a_{k}}$ in (\ref{f18}) possess ghost number $\left( k+1\right) $ and
Grassmann parity $\left( k+1\right) \mathrm{mod}\,2$, while those in (\ref
{f19}) have ghost number $\left( n-1\right) $ and Grassmann parity $\left(
n-1\right) \mathrm{mod}\,2$. The ghost number and Grassmann parity of the
antighosts follow from the general rules of the standard Hamiltonian BRST
formalism. The ghost number is defined in usual manner as the difference
between the pure ghost number (\textrm{pgh}) and the antighost number (%
\textrm{antigh}), where
\begin{equation}
\mathrm{pgh}\left( z^{A}\right) =0,\;\mathrm{pgh}\left( \eta ^{a_{0}}\right)
=1,\;\mathrm{pgh}\left( \mathcal{P}_{a_{0}}\right) =0,  \label{f23}
\end{equation}
\begin{equation}
\mathrm{pgh}\left( \eta ^{a_{k}}\right) =k+1,\;\mathrm{pgh}\left( \mathcal{P}%
_{a_{k}}\right) =0,\;k=1,\cdots ,n-3,  \label{f25}
\end{equation}
\begin{equation}
\mathrm{pgh}\left( \eta ^{a_{n-2}}\right) =n-1,\;\mathrm{pgh}\left( \mathcal{%
P}_{a_{n-2}}\right) =0,  \label{f27}
\end{equation}
\begin{equation}
\mathrm{antigh}\left( z^{A}\right) =0,\;\mathrm{antigh}\left( \eta
^{a_{0}}\right) =0,\;\mathrm{antigh}\left( \mathcal{P}_{a_{0}}\right) =1,
\label{f24}
\end{equation}
\begin{equation}
\mathrm{antigh}\left( \eta ^{a_{k}}\right) =0,\;\mathrm{antigh}\left(
\mathcal{P}_{a_{k}}\right) =k+1,\;k=1,\cdots ,n-3,  \label{f26}
\end{equation}
\begin{equation}
\mathrm{antigh}\left( \eta ^{a_{n-2}}\right) =0,\;\mathrm{antigh}\left(
\mathcal{P}_{a_{n-2}}\right) =n-1.  \label{f28}
\end{equation}

The BRST charge of this free model takes the form
\begin{eqnarray}
&&\Omega _{0}=\int d^{n-1}x\left( \eta ^{(1)a}G_{a}^{(1)}+\eta
^{a}G_{a}^{(2)}+\eta _{a}^{(1)ij}G_{ij}^{(1)a}\right.  \nonumber \\
&&+\eta _{a}^{ij}G_{ij}^{(2)a}+C_{i}^{(1)a}\gamma
_{a}^{(1)i}+C_{i}^{a}\gamma _{a}^{(2)i}  \nonumber \\
&&+\sum\limits_{k=3}^{n-1}\left( -\right) ^{k-1}\eta _{a}^{i_{1}i_{2}\cdots
i_{k}}\partial _{[i_{1}}\mathcal{P}_{i_{2}\cdots i_{k}]}^{a}  \nonumber \\
&&\left. +\sum\limits_{k=2}^{n-1}\left( -\right) ^{k-1}C_{i_{1}i_{2}\cdots
i_{k}}^{a}\partial ^{[i_{1}}P_{a}^{i_{2}\cdots i_{k}]}\right) ,  \label{f29}
\end{eqnarray}
while the corresponding BRST-invariant Hamiltonian, which is solution to the
equation (\ref{d8}), is expressed like
\begin{equation}
H_{0\mathrm{B}}=H_{0}+\int d^{n-1}x\left( \eta ^{(1)a}\mathcal{P}_{a}+\eta
_{a}^{(1)ij}\mathcal{P}_{ij}^{a}+C_{i}^{(1)a}P_{a}^{i}\right) .  \label{f30}
\end{equation}
In general, any function $F$ with $\mathrm{gh}\left( F\right) =0$ that
fulfills $\left[ F,\Omega _{0}\right] =0$ is called BRST observable. Since
the investigated theory has no physical degrees of freedom, its BRST
observables are BRST-exact, $F=\left[ M_{0},\Omega _{0}\right] $, for some $%
M_{0}$ with $\mathrm{gh}\left( M_{0}\right) =-1$. In consequence, the
BRST-invariant Hamiltonian will also be BRST-exact
\begin{equation}
H_{0\mathrm{B}}=\left[ K_{0},\Omega _{0}\right] ,  \label{f31}
\end{equation}
where
\begin{equation}
K_{0}=\int d^{n-1}x\left( H_{i}^{a}P_{a}^{i}-\frac{1}{2}B_{a}^{ij}\mathcal{P}%
_{ij}^{a}-A_{0}^{a}\mathcal{P}_{a}\right) .  \label{f32}
\end{equation}

The BRST charge encodes all the information on the gauge structure of the
first-class constraints. We remark that in our case the free BRST charge (%
\ref{f29}) breaks into terms with antighost numbers ranging from zero to $%
\left( n-2\right) $. The pieces with antighost number zero contain the
first-class constraint functions (\ref{f3}--\ref{f5}). If the algebra of the
first-class constraints is non-abelian, then there appear terms linear in
the antighost number one antighosts and quadratic in the pure ghost number
one ghosts. The absence of such terms in our case reflects that the
first-class constraints are abelian. The elements from (\ref{f29}) with
higher antighost number give us information on the reducibility functions (%
\ref{f15}--\ref{f16}). If the reducibility relations held on-shell, then
there would appear components linear in the ghosts for ghosts (ghosts of
pure ghost number strictly greater than one) and at least quadratic in the
various antighosts. Such pieces are not present in (\ref{f29}) since the
reducibility relations hold off-shell. Other possible components in the BRST
charge offer information on the higher-order structure functions related to
the first-class constraints. There are no such terms in (\ref{f29}), as a
consequence of the fact that all higher-order structure functions vanish for
the free theory analysed in the above. On the other hand, the BRST-invariant
Hamiltonian (\ref{f30}) decomposes into pieces of antighost number zero and
one. The element of antighost number zero is nothing but the first-class
Hamiltonian. The terms of antighost number one underlies the brackets
between the first-class Hamiltonian and the first-class constraints (the
relations (\ref{f12}--\ref{f14})). For a generic theory, there might appear
pieces of higher antighost number as well, that provide information on the
higher-order structure functions related to the first-class Hamiltonian. By
deforming the BRST charge and the BRST-invariant Hamiltonian, one deforms
everything, namely, the first-class constraints, their algebra, the
reducibility relations and their behaviour, the first-class Hamiltonian, its
brackets with the new first-class constraints, etc.

The BRST symmetry of the free theory, $s\cdot =\left[ \cdot ,\Omega
_{0}\right] $, splits as
\begin{equation}
s=\delta +\gamma ,  \label{f33}
\end{equation}
where $\delta $ denotes the Koszul-Tate differential ($\mathrm{antigh}\left(
\delta \right) =-1$, $\mathrm{pgh}\left( \delta \right) =0$), and $\gamma $
represents the exterior longitudinal derivative ($\mathrm{antigh}\left(
\gamma \right) =0$, $\mathrm{pgh}\left( \gamma \right) =1$). These two
operators act on the variables from the BRST complex like
\begin{equation}
\delta z^{A}=0,\;\delta \eta ^{a_{k}}=0,\;k=0,\cdots ,n-2,  \label{f34}
\end{equation}
\begin{equation}
\delta \mathcal{P}_{a}^{(1)}=-\pi _{a}^{0},\;\delta \mathcal{P}_{a}=\partial
_{i}B_{a}^{0i},\;\delta P_{a}^{(1)i}=p_{a}^{i},\;\delta P_{a}^{i}=-\partial
^{i}\varphi _{a},  \label{f35}
\end{equation}
\begin{equation}
\delta \mathcal{P}_{ij}^{(1)a}=-2\pi _{ij}^{a},\;\delta \mathcal{P}%
_{ij}^{a}=\partial _{[i}A_{j]}^{a},  \label{f36}
\end{equation}
\begin{equation}
\delta P_{a}^{i_{1}i_{2}\cdots i_{k}}=\left( -\right) ^{k}\partial
^{[i_{1}}P_{a}^{i_{2}\cdots i_{k}]},\;k=2,\cdots ,n-1,  \label{f37}
\end{equation}
\begin{equation}
\delta \mathcal{P}_{i_{1}i_{2}\cdots i_{k}}^{a}=\left( -\right) ^{k}\partial
_{[i_{1}}\mathcal{P}_{i_{2}\cdots i_{k}]}^{a},\;k=3,\cdots ,n-1,  \label{f38}
\end{equation}
\begin{equation}
\gamma A_{i}^{a}=\partial _{i}\eta ^{a},\;\gamma A_{0}^{a}=\eta
^{(1)a},\;\gamma \varphi _{a}=0,\;\gamma \pi _{a}^{0}=0,\;\gamma
p_{a}^{i}=0,\;\gamma \pi _{ij}^{a}=0,  \label{f39}
\end{equation}
\begin{equation}
\gamma B_{a}^{0i}=2\partial _{j}\eta _{a}^{ij},\;\gamma B_{a}^{ij}=2\eta
_{a}^{(1)ij},\;\gamma H_{i}^{a}=-C_{i}^{(1)a},\;\gamma H_{0}^{a}=\partial
^{i}C_{i}^{a},  \label{f40}
\end{equation}
\begin{equation}
\gamma \eta ^{(1)a}=\gamma \eta ^{a}=\gamma C_{i}^{(1)a}=\gamma \eta
_{a}^{(1)ij}=0,  \label{f41}
\end{equation}
\begin{equation}
\gamma \eta _{a}^{ij}=3\partial _{k}\eta _{a}^{ijk},\;\gamma
C_{i}^{a}=2\partial ^{j}C_{ij}^{a},  \label{f42}
\end{equation}
\begin{equation}
\gamma \eta _{a}^{i_{1}\cdots i_{k}}=\left( k+1\right) \partial _{i}\eta
_{a}^{ii_{1}\cdots i_{k}},\;k=3,\cdots ,n-2,  \label{f43}
\end{equation}
\begin{equation}
\gamma C_{i_{1}\cdots i_{k}}^{a}=-\left( k+1\right) \partial
^{i}C_{ii_{1}\cdots i_{k}}^{a},\;k=2,\cdots ,n-2,  \label{f44}
\end{equation}
\begin{equation}
\gamma \eta _{a}^{i_{1}\cdots i_{n-1}}=0,\;\gamma C_{i_{1}\cdots
i_{n-1}}^{a}=0,  \label{f45}
\end{equation}
\begin{equation}
\gamma \mathcal{P}_{a_{k}}=0,\;k=0,\cdots ,n-2.  \label{f46}
\end{equation}
The last formulas will be employed in the next section at the deformation of
the free theory.

\section{Deformation of the BRST charge}

In this section we solve the equations (\ref{d4}--\ref{d5}), etc.,
that govern the deformation of the BRST charge in the case of the
free model under study by relying on cohomological techniques. We
will focus only on both local and spacetime-dimension independent
deformations. As a result, we find that only the first-order
deformation is non-trivial, while its consistency is equivalent to
the existence of a `two-tensor' (in the collection indices)
depending on the undifferentiated scalar fields, that must be
antisymmetric and fulfill a certain identity.

\subsection{First-order deformation}

Initially, we solve the equation (\ref{d4}), which is responsible for the
first-order deformation of the BRST charge. Using the notations from (\ref
{d1}), it takes the local form
\begin{equation}
s\omega _{1}=\partial _{i}j^{i},  \label{c1}
\end{equation}
for some local $j^{i}$. In order to investigate this equation, we develop $%
\omega _{1}$ according to the antighost number and suppose that the
development stops at a finite order
\begin{equation}
\omega _{1}=\stackrel{(0)}{\omega }_{1}+\stackrel{(1)}{\omega }_{1}+\cdots +%
\stackrel{(J)}{\omega }_{1},\;\mathrm{antigh}\left( \stackrel{(I)}{\omega }%
_{1}\right) =I,\;\mathrm{gh}\left( \stackrel{(I)}{\omega }_{1}\right) =1,
\label{c2}
\end{equation}
where the last term can be assumed to be annihilated by $\gamma $,
\begin{equation}
\gamma \stackrel{(J)}{\omega }_{1}=0.  \label{gama}
\end{equation}
Thus, we need to know the cohomology of $\gamma $, $H\left( \gamma \right) $%
, in order to determine the piece of highest antighost number in
(\ref{c2}). From the actions (\ref{f39}--\ref{f46}) of $\gamma $
acting on the BRST generators of the BRST complex, we remark that
$H\left( \gamma \right) $ is generated by
\begin{equation}
\Phi ^{\alpha }=\left( F_{ij}^{a}=\partial _{[i}A_{j]}^{a},\varphi _{a},\pi
_{a}^{0},p_{a}^{i},\pi _{ij}^{a},\partial _{i}B_{a}^{0i}\right) ,  \label{c3}
\end{equation}
(together with their spatial derivatives up to a finite order), by
the antighosts (\ref{f20}--\ref{f22}) and their spatial
derivatives up to a finite order, as well as by the
undifferentiated ghosts $\eta ^{a}$, $\eta
_{a}^{i_{1}\cdots i_{n-1}}$ and $C_{i_{1}\cdots i_{n-1}}^{a}$. (The ghosts $%
\eta ^{(1)a}$, $C_{i}^{(1)a}$ and $\eta _{a}^{(1)ij}$, although $\gamma $%
-invariant, are also $\gamma $-exact, and hence trivial in $H\left( \gamma
\right) $. Same with respect to the spatial part of the spacetime
derivatives of $\eta ^{a}$, $\eta _{a}^{i_{1}\cdots i_{n-1}}$ and $%
C_{i_{1}\cdots i_{n-1}}^{a}$.) In this way, the general solution to the
equation (\ref{gama}) can be written (up to a trivial term) as
\begin{equation}
\stackrel{(J)}{\omega }_{1}=a_{J}\left( \left[ \Phi ^{\alpha }\right]
,\left[ \mathcal{P}_{a_{k}}\right] _{k=0,\cdots ,n-2}\right) e^{J+1}\left(
\eta ^{a},\eta _{a}^{i_{1}\cdots i_{n-1}},C_{i_{1}\cdots i_{n-1}}^{a}\right)
,  \label{c5}
\end{equation}
where $e^{J+1}\left( \eta ^{a},\eta _{a}^{i_{1}\cdots
i_{n-1}},C_{i_{1}\cdots i_{n-1}}^{a}\right) $ stand for the elements with
pure ghost number equal to $\left( J+1\right) $ of a basis in the space of
the polynomials in the corresponding ghosts, and obviously $\mathrm{antigh}%
\left( a_{J}\right) =J$. The notation $f\left( \left[ q\right] \right) $
signifies that $f$ depends on $q$ and its spatial derivatives up to a finite
order.

The equation (\ref{c1}) projected on antighost number $\left( J-1\right) $
becomes
\begin{equation}
\delta \stackrel{(J)}{\omega }_{1}+\gamma \stackrel{(J-1)}{\omega }%
_{1}=\partial ^{i}\stackrel{(J)}{m}_{i},  \label{c6}
\end{equation}
and it shows that a necessary condition for the existence of $\stackrel{(J-1)%
}{\omega }_{1}$ is that the functions $a_{J}$ from (\ref{c5}) belong to $%
H_{J}\left( \delta |\tilde{d}\right) $, where the last notation means the
homological space of the Koszul-Tate differential modulo the spatial part of
the exterior spacetime derivative at antighost number $J$. Equivalently,
these functions should satisfy the equation
\begin{equation}
\delta a_{J}=\partial ^{i}n_{i},\;\mathrm{antigh}\left( n_{i}\right) =J-1.
\label{c6aa}
\end{equation}
Translating the Lagrangian results from \cite{gen} at the Hamiltonian level,
as our model is $\left( n-2\right) $-order reducible and the constraint
functions are linear in the reduced phase-space variables, we have that
\begin{equation}
H_{K}\left( \delta |\tilde{d}\right) =0,\;\mathrm{for}\;K>n-1.  \label{c7}
\end{equation}

Consequently, we can assume that $J=n-1$, and thus the development (\ref{c2}%
) stops after the first $n$ terms
\begin{equation}
\omega _{1}=\stackrel{(0)}{\omega }_{1}+\stackrel{(1)}{\omega }_{1}+\cdots +%
\stackrel{(n-1)}{\omega }_{1},  \label{c8}
\end{equation}
with $\stackrel{(n-1)}{\omega }_{1}$ given by (\ref{c5}) with $J=n-1$ and $%
a_{n-1}$ from $H_{n-1}\left( \delta |\tilde{d}\right) $. After some
computation, we find that the most general representative of $H_{n-1}\left(
\delta |\tilde{d}\right) $ is expressed like
\begin{eqnarray}
&&a_{n-1}^{i_{1}\cdots i_{n-1}}=\frac{\delta U}{\delta \varphi _{a}}%
P_{a}^{i_{1}\cdots i_{n-1}}+\sum\limits_{p=2}^{n-1}\sum\limits_{1\leq
j_{1}\leq j_{2}\leq \cdots \leq j_{p}<n-1}\frac{\delta ^{p}U}{\delta \varphi
_{a_{1}}\delta \varphi _{a_{2}}\cdots \delta \varphi _{a_{p}}}\times
\nonumber \\
&&\times P_{a_{1}}^{[i_{1}\cdots i_{j_{1}}}P_{a_{2}}^{i_{j_{1}+1}\cdots
i_{j_{1}+j_{2}}}\cdots P_{a_{p-1}}^{i_{j_{1}+\cdots +j_{p-2}+1}\cdots
i_{j_{1}+\cdots +j_{p-1}}}P_{a_{p}}^{i_{j_{1}+\cdots +j_{p-1}+1}\cdots
i_{n-1}]},  \label{c9}
\end{eqnarray}
where $U$ is an arbitrary function involving the undifferentiated scalar
fields $\varphi _{a}$ and
\begin{equation}
j_{p}=n-1-\left( j_{1}+j_{2}+\cdots +j_{p-1}\right) .  \label{notlong}
\end{equation}
Now, we can completely determine the last component in (\ref{c8}). The
elements of $e^{n}\left( \eta ^{a},\eta _{a}^{i_{1}\cdots
i_{n-1}},C_{i_{1}\cdots i_{n-1}}^{a}\right) $ can be written in the form
\begin{equation}
\eta ^{a}C_{i_{1}\cdots i_{n-1}}^{b},\eta ^{a}\eta ^{b}\eta
_{c}^{i_{1}\cdots i_{n-1}},\eta ^{a_{1}}\eta ^{a_{2}}\cdots \eta ^{a_{n}},
\label{c13}
\end{equation}
for $n\geq 4$\footnote{%
For $n=4$ there is an extra possibility because $\eta _{a}^{i_{1}\cdots
i_{n-1}}\rightarrow \eta _{a}^{ijk}$, with $pgh\left( \eta _{a}^{ijk}\right)
=2$, and so we have a supplementary element of the basis in the ghosts at
pure ghost number $n=4$, namely, $\eta _{a}^{ijk}\eta _{b}^{i^{\prime
}j^{\prime }k^{\prime }}$. However, this element can be discarded \cite
{ijmpa}, so finally (\ref{c13}) still covers all the investigated situations.%
}. It means that the piece of highest antighost number in the first-order
deformation is determined once we `glue' (\ref{c9}) to (\ref{c13}) like in (%
\ref{c5}). The last component in (\ref{c13}) needs the adjustment
of a completely antisymmetric constant $K_{i_{1}\cdots i_{n-1}}$
in order to match (\ref{c9}), which can only be by `covariance'
arguments proportional to the spatial part of the completely
antisymmetric symbol in $n$ dimensions, $\varepsilon
_{0i_{1}\cdots i_{n-1}}$. As mentioned in the above, we ask that
the resulting deformations are independent of the spacetime
dimension. If we add to the last component
$\stackrel{(n-1)}{\omega }_{1}$ a term involving the completely
antisymmetric symbol, this element will generate in the deformed
BRST-invariant Hamiltonian (by means of the equations
(\ref{d9}--\ref{d10}), etc.) some pieces that contain this symbol,
such that the Lagrangian action of the interacting model will
accordingly exhibit some vertices that break the PT-invariance. By
imposing the PT-invariance at the level of the coupled gauge
theory, the third element in (\ref{c13}) should be removed. Then,
we finally obtain that
\begin{equation}
\stackrel{(n-1)}{\omega }_{1}=-W_{ab}^{i_{1}\cdots i_{n-1}}\eta
^{a}C_{i_{1}\cdots i_{n-1}}^{b}-\frac{\left( -\right) ^{n}}{2}\left(
M_{ab}^{c}\right) ^{i_{1}\cdots i_{n-1}}\eta ^{a}\eta ^{b}\eta
_{ci_{1}\cdots i_{n-1}},  \label{c14}
\end{equation}
where $W_{ab}^{i_{1}\cdots i_{n-1}}$ and $\left( M_{ab}^{c}\right)
^{i_{1}\cdots i_{n-1}}$ are obtained from $a_{n-1}^{i_{1}\cdots i_{n-1}}$
with the function $U$ on the scalar fields replaced by a `two-tensor' $%
W_{ab} $, respectively, a `two--one-tensor' $M_{ab}^{c}$%
\begin{eqnarray}
&&W_{ab}^{i_{1}\cdots i_{n-1}}=\frac{\delta W_{ab}}{\delta \varphi _{a}}%
P_{a}^{i_{1}\cdots i_{n-1}}+\sum\limits_{p=2}^{n-1}\sum\limits_{1\leq
j_{1}\leq j_{2}\leq \cdots \leq j_{p}<n-1}\frac{\delta ^{p}W_{ab}}{\delta
\varphi _{a_{1}}\delta \varphi _{a_{2}}\cdots \delta \varphi _{a_{p}}}\times
\nonumber \\
&&\times P_{a_{1}}^{[i_{1}\cdots i_{j_{1}}}P_{a_{2}}^{i_{j_{1}+1}\cdots
i_{j_{1}+j_{2}}}\cdots P_{a_{p-1}}^{i_{j_{1}+\cdots +j_{p-2}+1}\cdots
i_{j_{1}+\cdots +j_{p-1}}}P_{a_{p}}^{i_{j_{1}+\cdots +j_{p-1}+1}\cdots
i_{n-1}]},  \label{c14a}
\end{eqnarray}
\begin{eqnarray}
&&\left( M_{ab}^{c}\right) ^{i_{1}\cdots i_{n-1}}= \frac{\delta M_{ab}^{c}}{%
\delta \varphi _{a}}P_{a}^{i_{1}\cdots
i_{n-1}}+\sum\limits_{p=2}^{n-1}\sum\limits_{1\leq j_{1}\leq j_{2}\leq
\cdots \leq j_{p}<n-1}\frac{\delta ^{p}M_{ab}^{c}}{\delta \varphi
_{a_{1}}\delta \varphi _{a_{2}}\cdots \delta \varphi _{a_{p}}}\times
\nonumber \\
&&\times P_{a_{1}}^{[i_{1}\cdots i_{j_{1}}}P_{a_{2}}^{i_{j_{1}+1}\cdots
i_{j_{1}+j_{2}}}\cdots P_{a_{p-1}}^{i_{j_{1}+\cdots +j_{p-2}+1}\cdots
i_{j_{1}+\cdots +j_{p-1}}}P_{a_{p}}^{i_{j_{1}+\cdots +j_{p-1}+1}\cdots
i_{n-1}]}.  \label{c14b}
\end{eqnarray}
The notion of `tensor' has no other significance for the moment than to
emphasise that these functions in the dynamical fields $\varphi _{a}$ carry
more than one collection index differently positioned. Moreover, $M_{ab}^{c}$
is antisymmetric in its lower indices due to the anticommutation of the
fermionic ghosts $\eta ^{a}$. The additional constants in (\ref{c14}) were
introduced for convenience.

Taking into account the actions (\ref{f34}--\ref{f38}) of the Koszul-Tate
differential, we can prove the recursive relations
\begin{equation}
\delta a_{k}^{i_{1}i_{2}\cdots i_{k}}=\left( -\right) ^{k}\partial
^{[i_{1}}a_{k-1}^{i_{2}\cdots i_{k}]},\;k=1,\cdots ,n-1,  \label{c10}
\end{equation}
where for $k=2,\cdots ,n-2$ we have
\begin{eqnarray}
&&a_{k}^{i_{1}\cdots i_{k}}=\frac{\delta U}{\delta \varphi _{a}}%
P_{a}^{i_{1}\cdots i_{k}}+\sum\limits_{q=2}^{k}\sum\limits_{1\leq j_{1}\leq
j_{2}\leq \cdots \leq j_{q}<k}\frac{\delta ^{q}U}{\delta \varphi
_{a_{1}}\delta \varphi _{a_{2}}\cdots \delta \varphi _{a_{q}}}\times
\nonumber \\
&&\times P_{a_{1}}^{[i_{1}\cdots i_{j_{1}}}P_{a_{2}}^{i_{j_{1}+1}\cdots
i_{j_{1}+j_{2}}}\cdots P_{a_{q-1}}^{i_{j_{1}+\cdots +j_{q-2}+1}\cdots
i_{j_{1}+\cdots +j_{q-1}}}P_{a_{q}}^{i_{j_{1}+\cdots +j_{q-1}+1}\cdots
i_{k}]},  \label{c11}
\end{eqnarray}
and
\begin{equation}
a_{1}^{i}=\frac{\delta U}{\delta \varphi _{a}}P_{a}^{i},\;a_{0}=U.
\label{c12}
\end{equation}
In (\ref{c11}) we used the notation $j_{q}=k-\left( j_{1}+j_{2}+\cdots
+j_{q-1}\right) $. The formulas (\ref{c10}--\ref{c12}) hold in general, for
any starting $a_{n-1}^{i_{1}\cdots i_{n-1}}$ of the form (\ref{c9}), such
that they are also valid for $U$ replaced with $W_{ab}$, respectively, $%
M_{ab}^{c}$, like in (\ref{c14a}--\ref{c14b}). In the sequel, we will
constantly use the above relations in order to infer the components of lower
antighost number in $\Omega _{1}$. Thus, the piece of antighost number $%
\left( n-2\right) $ from the first-order deformation, which is solution to
the equation (\ref{c6}) with $J=n-1$, reads as
\begin{eqnarray}
&&\stackrel{(n-2)}{\omega }_{1}=-W_{ab}^{i_{1}\cdots i_{n-2}}\eta
^{a}C_{i_{1}\cdots i_{n-2}}^{b}+\frac{\left( -\right) ^{n}}{2}\left(
M_{ab}^{c}\right) ^{i_{1}\cdots i_{n-2}}\eta ^{a}\eta ^{b}\eta
_{ci_{1}\cdots i_{n-2}}  \nonumber \\
&&-C_{n-1}^{1}W_{ab}^{i_{1}\cdots i_{n-2}}A^{ai_{n-1}}C_{i_{1}\cdots
i_{n-1}}^{b}  \nonumber \\
&&-\sum\limits_{k=3}^{n}\left( -\right) ^{k}C_{n-1}^{k-1}W_{ab}^{i_{1}\cdots
i_{n-k}}\mathcal{P}^{ai_{n-k+1}\cdots i_{n-1}}C_{i_{1}\cdots i_{n-1}}^{b}
\nonumber \\
&&-\left( -\right) ^{n}C_{n-1}^{1}\left( M_{ab}^{c}\right) ^{i_{1}\cdots
i_{n-2}}A^{ai_{n-1}}\eta ^{b}\eta _{ci_{1}\cdots i_{n-1}}  \nonumber \\
&&-\left( -\right) ^{n}\sum\limits_{k=3}^{n}\left( -\right)
^{k}C_{n-1}^{k-1}\left( M_{ab}^{c}\right) ^{i_{1}\cdots i_{n-k}}\mathcal{P}%
^{ai_{n-k+1}\cdots i_{n-1}}\eta ^{b}\eta _{ci_{1}\cdots i_{n-1}}.
\label{c15}
\end{eqnarray}
The equation (\ref{c1}) projected on antighost number $\left( n-3\right) $
becomes precisely the equation (\ref{c6}) for $J=n-2$, which further yields
\begin{eqnarray}
&&\stackrel{(n-3)}{\omega }_{1}=-W_{ab}^{i_{1}\cdots i_{n-3}}\eta
^{a}C_{i_{1}\cdots i_{n-3}}^{b}  \nonumber \\
&&-\frac{\left( -\right) ^{n}}{2}\left( M_{ab}^{c}\right) ^{i_{1}\cdots
i_{n-3}}\eta ^{a}\eta ^{b}\eta _{ci_{1}\cdots i_{n-3}}  \nonumber \\
&&-C_{n-2}^{1}W_{ab}^{i_{1}\cdots i_{n-3}}A^{ai_{n-2}}C_{i_{1}\cdots
i_{n-2}}^{b}  \nonumber \\
&&+\sum\limits_{k=4}^{n}\left( -\right) ^{k}C_{n-2}^{k-2}W_{ab}^{i_{1}\cdots
i_{n-k}}\mathcal{P}^{ai_{n-k+1}\cdots i_{n-2}}C_{i_{1}\cdots i_{n-2}}^{b}
\nonumber \\
&&+\left( -\right) ^{n}C_{n-2}^{1}\left( M_{ab}^{c}\right) ^{i_{1}\cdots
i_{n-3}}A^{ai_{n-2}}\eta ^{b}\eta _{ci_{1}\cdots i_{n-2}}  \nonumber \\
&&-\left( -\right) ^{n}\sum\limits_{k=4}^{n}\left( -\right)
^{k}C_{n-2}^{k-2}\left( M_{ab}^{c}\right) ^{i_{1}\cdots i_{n-k}}\mathcal{P}%
^{ai_{n-k+1}\cdots i_{n-2}}\eta ^{b}\eta _{ci_{1}\cdots i_{n-2}}  \nonumber
\\
&&-\left( -\right) ^{n}\sum\limits_{p=2}^{\left[ \frac{n-2}{2}\right]
}\sum\limits_{q=p+1}^{n-p-1}\left( -\right)
^{q}C_{n-1}^{p}C_{n-p-1}^{q}\left( M_{ab}^{c}\right) ^{i_{1}\cdots
i_{n-p-q-1}}\times  \nonumber \\
&&\times \mathcal{P}^{aj_{1}\cdots j_{q}}\mathcal{P}^{bl_{1}\cdots
l_{p}}\eta _{ci_{1}\cdots i_{n-p-q-1}j_{1}\cdots j_{q}l_{1}\cdots l_{p}}
\nonumber \\
&&-\frac{\left( -\right) ^{n}}{2}\sum\limits_{k=2}^{\left[ \frac{n-1}{2}%
\right] }\left( -\right) ^{k}C_{n-1}^{k}C_{n-k-1}^{k}\left(
M_{ab}^{c}\right) ^{i_{1}\cdots i_{n-2k-1}}\times  \nonumber \\
&&\mathcal{P}^{aj_{1}\cdots j_{k}}\mathcal{P}^{bl_{1}\cdots l_{k}}\eta
_{ci_{1}\cdots i_{n-2k-1}j_{1}\cdots j_{k}l_{1}\cdots l_{k}}  \nonumber \\
&&+\left( -\right) ^{n}C_{n-1}^{2}\left( M_{ab}^{c}\right) ^{i_{1}\cdots
i_{n-3}}A^{ai_{n-2}}A^{bi_{n-1}}\eta _{ci_{1}i_{2}\cdots i_{n-1}}.
\label{c16}
\end{eqnarray}
In the same manner, we can solve the equation (\ref{c1}) for any antighost
number $\left( n-m\right) $, where $m=4,\cdots ,n-2$, and derive the
solutions
\begin{eqnarray}
&&\stackrel{(n-m)}{\omega }_{1}=-W_{ab}^{i_{1}\cdots i_{n-m}}\eta
^{a}C_{i_{1}\cdots i_{n-m}}^{b}  \nonumber \\
&&-C_{n-m+1}^{1}W_{ab}^{i_{1}\cdots i_{n-m}}A^{ai_{n-m+1}}C_{i_{1}\cdots
i_{n-m+1}}^{b}  \nonumber \\
&&-\sum\limits_{k=m+1}^{n}\left( -\right)
^{k+m}C_{n-m+1}^{k-m+1}W_{ab}^{i_{1}\cdots i_{n-k}}\mathcal{P}%
^{ai_{n-k+1}\cdots i_{n-m+1}}C_{i_{1}\cdots i_{n-m+1}}^{b}  \nonumber \\
&&+\frac{\left( -\right) ^{m+n}}{2}\left( M_{ab}^{c}\right) ^{i_{1}\cdots
i_{n-m}}\eta ^{a}\eta ^{b}\eta _{ci_{1}\cdots i_{n-m}}  \nonumber \\
&&-\left( -\right) ^{m+n}C_{n-m+1}^{1}\left( M_{ab}^{c}\right) ^{i_{1}\cdots
i_{n-m}}A^{ai_{n-m+1}}\eta ^{b}\eta _{ci_{1}\cdots i_{n-m+1}}  \nonumber \\
&&-\sum\limits_{k=m+1}^{n}\left( -\right) ^{k+n}C_{n-m+1}^{k-m+1}\left(
M_{ab}^{c}\right) ^{i_{1}\cdots i_{n-k}}\mathcal{P}^{ai_{n-k+1}\cdots
i_{n-m+1}}\eta ^{b}\eta _{ci_{1}\cdots i_{n-m+1}}  \nonumber \\
&&-\left( -\right) ^{m+n}C_{n-m+2}^{2}\left( M_{ab}^{c}\right) ^{i_{1}\cdots
i_{n-m}}A^{ai_{n-m+1}}A^{bi_{n-m+2}}\eta _{ci_{1}\cdots i_{n-m+2}}  \nonumber
\\
&&+\frac{1}{2}\sum\limits_{k=2}^{\left[ \frac{n-m+2}{2}\right] }\left(
-\right) ^{k+n+m}C_{n-m+2}^{k}C_{n-m-k+2}^{k}\left( M_{ab}^{c}\right)
^{i_{1}\cdots i_{n-2k-m+2}}\times  \nonumber \\
&&\times \mathcal{P}^{aj_{1}\cdots j_{k}}\mathcal{P}^{bl_{1}\cdots
l_{k}}\eta _{ci_{1}\cdots i_{n-2k-m+2}j_{1}\cdots j_{k}l_{1}\cdots l_{k}}
\nonumber \\
&&+\sum\limits_{p=2}^{\left[ \frac{n-m+1}{2}\right]
}\sum\limits_{q=p+1}^{n-m-p+2}\left( -\right)
^{q+n+m}C_{n-m+2}^{p}C_{n-m-p+2}^{q}\left( M_{ab}^{c}\right) ^{i_{1}\cdots
i_{n-m-p-q+2}}\times  \nonumber \\
&&\times \mathcal{P}^{aj_{1}\cdots j_{q}}\mathcal{P}^{bl_{1}\cdots
l_{p}}\eta _{ci_{1}\cdots i_{n-m-p-q+2}j_{1}\cdots j_{q}l_{1}\cdots l_{p}}
\nonumber \\
&&-\sum\limits_{k=m+1}^{n}\left( -\right)
^{k+n}C_{n-m+1}^{k-m+1}C_{n-m+2}^{1}\left( M_{ab}^{c}\right) ^{i_{1}\cdots
i_{n-k}}\times  \nonumber \\
&&\times \mathcal{P}^{ai_{n-k+1}\cdots i_{n-m+1}}A^{bi_{n-m+2}}\eta
_{ci_{1}i_{2}\cdots i_{n-m+2}}.  \label{c17}
\end{eqnarray}
The equation (\ref{c1}) projected on antighost number one becomes exactly
the equation (\ref{c6}) for $J=2$, which leads to
\begin{eqnarray}
&&\stackrel{(1)}{\omega }_{1}=-\frac{\delta W_{ab}}{\delta \varphi _{c}}%
P_{c}^{i}\left( \eta ^{a}C_{i}^{b}+2A^{aj}C_{ij}^{b}\right) +W_{ab}\mathcal{P%
}^{aij}C_{ij}^{b}  \nonumber \\
&&+\frac{\delta M_{ab}^{c}}{\delta \varphi _{d}}P_{di}\left( \frac{1}{2}\eta
^{a}\eta ^{b}B_{c}^{0i}+2A_{j}^{a}\eta ^{b}\eta
_{c}^{ij}+3A_{j}^{a}A_{k}^{b}\eta _{c}^{ijk}\right)  \nonumber \\
&&-M_{ab}^{c}\left( \mathcal{P}_{ij}^{a}\eta ^{b}\eta _{c}^{ij}+\mathcal{P}%
_{[ij}^{a}A_{k]}^{b}\eta _{c}^{ijk}+\frac{1}{2}\eta ^{a}\eta ^{b}\mathcal{P}%
_{c}\right) .  \label{c18}
\end{eqnarray}
Finally, using the equation (\ref{c6}) for $J=1$, we generate $\stackrel{(0)%
}{\omega }_{1}$ under the form
\begin{equation}
\stackrel{(0)}{\omega }_{1}=W_{ab}\left( \eta
^{a}H_{0}^{b}-A^{ai}C_{i}^{b}\right) +M_{ab}^{c}\left( A_{i}^{a}\eta
^{b}B_{c}^{0i}-A_{i}^{a}A_{j}^{b}\eta _{c}^{ij}\right) .  \label{c19}
\end{equation}
In consequence, we succeeded in finding the complete form of the first-order
deformation of the BRST charge for the model under study, which reduces to
the sum among the right hand-sides of the formulas (\ref{c14}) and (\ref{c15}%
--\ref{c19}).

\subsection{Higher-order deformations}

The next target is to investigate the consistency of the first-order
deformation of the BRST charge, described by the equation (\ref{d5}). By
direct computation, it follows that $\left[ \Omega _{1},\Omega _{1}\right]
=\int d^{n-1}x\,\Delta $, with
\begin{eqnarray}
&&\Delta =K^{abc}t_{abc}+\sum\limits_{k=1}^{n-1}K_{a_{1}\cdots a_{k}}^{abc}%
\frac{\delta ^{k}t_{abc}}{\delta \varphi _{a_{1}}\cdots \delta \varphi
_{a_{k}}}  \nonumber \\
&&+K_{d,}^{abc}t_{abc}^{d}+\sum\limits_{k=1}^{n-1}K_{d,a_{1}\cdots
a_{k}}^{abc}\frac{\delta ^{k}t_{abc}^{d}}{\delta \varphi _{a_{1}}\cdots
\delta \varphi _{a_{k}}},  \label{c20}
\end{eqnarray}
where we made the notations
\begin{equation}
t_{abc}=W_{ec}M_{ab}^{e}+W_{ea}\frac{\delta W_{bc}}{\delta \varphi _{e}}%
+W_{eb}\frac{\delta W_{ca}}{\delta \varphi _{e}},  \label{c21}
\end{equation}
\begin{equation}
t_{abc}^{d}=W_{e[a}\frac{\delta M_{bc]}^{d}}{\delta \varphi _{e}}%
+M_{e[a}^{d}M_{bc]}^{e}.  \label{c22}
\end{equation}
On the one hand, the objects $K^{abc}$, $K_{d,}^{abc}$, $K_{a_{1}\cdots
a_{k}}^{abc}$ and $K_{d,a_{1}\cdots a_{k}}^{abc}$ are polynomials of ghost
number two involving the (undifferentiated) ghosts, antighosts, and fields $%
B_{a}^{0i}$ and $A_{i}^{a}$, such that they are not BRST-exact. For
instance, the terms corresponding to $k=n-1$ in (\ref{c20}) have the
concrete form
\begin{equation}
K_{a_{1}\cdots a_{n-1}}^{abc}=\left( -\right) ^{n}P_{a_{1}}^{i_{1}}\cdots
P_{a_{n-1}}^{i_{n-1}}\eta ^{a}\eta ^{b}C_{i_{1}\cdots i_{n-1}}^{c},
\label{c23}
\end{equation}
\begin{equation}
K_{d,a_{1}\cdots a_{n-1}}^{abc}=\frac{1}{3}P_{a_{1}}^{i_{1}}\cdots
P_{a_{n-1}}^{i_{n-1}}\eta ^{a}\eta ^{b}\eta ^{c}\eta _{di_{1}\cdots i_{n-1}}.
\label{c24}
\end{equation}
The general form of the functions $K^{abc}$, $K_{d,}^{abc}$, $K_{a_{1}\cdots
a_{k}}^{abc}$ and $K_{d,a_{1}\cdots a_{k}}^{abc}$ is complicated and not
illuminating for subsequent purposes. On the other hand, the equation (\ref
{d5}) requires that $\left[ \Omega _{1},\Omega _{1}\right] $ is $s$-exact.
However, since none of the terms in (\ref{c20}) is so, $\Delta $ must
vanish. This takes place if and only if the following equations are
simultaneously obeyed
\begin{equation}
t_{abc}=0,\;t_{abc}^{d}=0.  \label{c25}
\end{equation}
Analysing the structure of (\ref{c21}--\ref{c22}), we reach the conclusion
that the solution to (\ref{c25}) reads as
\begin{equation}
M_{ab}^{c}=\frac{\delta W_{ab}}{\delta \varphi _{c}},  \label{c26}
\end{equation}
where, in addition, the now antisymmetric `two-tensor' $W_{ab}$ is imposed
to fulfill the identity
\begin{equation}
W_{e[a}\frac{\delta W_{bc]}}{\delta \varphi _{e}}=0\Leftrightarrow W_{ea}%
\frac{\delta W_{bc}}{\delta \varphi _{e}}+\mathrm{cyclic}\left( a,b,c\right)
=0.  \label{c27}
\end{equation}
Under these conditions, we can further take $\Omega _{2}=0$, the remaining
higher-order deformation equations being satisfied with the choice
\begin{equation}
\Omega _{k}=0,\;k>2.  \label{c28}
\end{equation}

In consequence, the consistency of the first-order deformation of the BRST
charge for the free model under discussion implements two types of
conditions. First, it restricts that $W_{ab}$ and $M_{ab}^{c}$ are no longer
independent, but related like in (\ref{c26}). This immediately forces the
antisymmetry of $W_{ab}\left( \varphi _{a}\right) $ with respect to its
collection indices since $M_{ab}^{c}$ was already antisymmetric. Second, the
antisymmetric `two-tensor' is constrained to verify the identity (\ref{c27}%
). (We will comment more on the interpretation of these results at the end
of Section 6.) In this way the complete deformation of the BRST charge,
consistent to all orders in the coupling constant, reduces to
\begin{equation}
\Omega =\Omega _{0}+g\Omega _{1}=\Omega _{0}+g\int
d^{n-1}x\sum\limits_{k=0}^{n-1}\stackrel{(k)}{\omega }_{1},  \label{c29}
\end{equation}
where $\stackrel{(k)}{\omega }_{1}$ are read from (\ref{c14}) and (\ref{c15}%
--\ref{c19}), with $M_{ab}^{c}=\delta W_{ab}/\delta \varphi _{c}$.

\section{Deformation of the BRST-invariant Hamiltonian}

We now turn our attention to the BRST-invariant Hamiltonian
(\ref{f30}), whose deformation is stipulated by the equations
(\ref{d9}--\ref{d10}), etc. Like in the previous section, we
investigate only local deformations. Initially, we approach the
equation (\ref{d9}) associated with its first-order deformation.
Inserting (\ref{f31}) in (\ref{d9}) and using (\ref {d4}), on
behalf of Jacobi's identity for the Dirac bracket, we find the
equation $\left[ H_{1}-\left[ K_{0},\Omega _{1}\right] ,\Omega
_{0}\right] =0 $, showing that $H_{1}-\left[ K_{0},\Omega
_{1}\right] $ is a BRST observable of the free theory. As
mentioned in Section 3, all BRST observables are also BRST-exact,
or, in other words, they belong to the same equivalence class as
the trivial observable zero. In consequence, we can take
\begin{equation}
H_{1}=\left[ K_{0},\Omega _{1}\right] ,  \label{h1}
\end{equation}
where the function $K_{0}$ is displayed in (\ref{f32}). The expression (\ref
{h1}) offers us the first-order deformation of the BRST-invariant
Hamiltonian like
\begin{eqnarray}
&&h_{1}=-W_{ab}H_{\mu }^{a}A^{b\mu }-\frac{1}{2}M_{ab}^{c}A_{\mu }^{a}A_{\nu
}^{b}B_{c}^{\mu \nu }  \nonumber \\
&&-M_{ab}^{c}\left( \frac{1}{2}B_{c}^{ij}\eta ^{a}\mathcal{P}%
_{ij}^{b}+A_{0}^{a}\mathcal{P}_{ij}^{b}\eta _{c}^{ij}+A_{0}^{a}\eta ^{b}%
\mathcal{P}_{c}\right)  \nonumber \\
&&+\frac{\delta W_{ab}}{\delta \varphi _{c}}P_{c}^{i}\left( H_{i}^{a}\eta
^{b}+C_{i}^{a}A_{0}^{b}\right)  \nonumber \\
&&+\frac{\delta M_{ab}^{c}}{\delta \varphi _{d}}P_{di}\left( \eta
^{a}A_{j}^{b}B_{c}^{ij}-\eta ^{a}A_{0}^{b}B_{c}^{0i}+2A_{0}^{a}A_{j}^{b}\eta
_{c}^{ij}\right)  \nonumber \\
&&+\frac{1}{4}\left( \frac{\delta M_{ab}^{c}}{\delta \varphi _{d}}P_{dij}+%
\frac{\delta ^{2}M_{ab}^{c}}{\delta \varphi _{d}\delta \varphi _{e}}%
P_{di}P_{ej}\right) \eta ^{a}\eta ^{b}B_{c}^{ij}  \nonumber \\
&&+\sum\limits_{k=2}^{n-1}A_{0}^{a}\frac{\partial ^{L}\stackrel{(k)}{\omega }%
_{1}}{\partial \eta ^{a}},  \label{h2}
\end{eqnarray}
where we made the notation $\partial ^{L}/\partial \eta ^{a}$ for the left
derivative with respect to $\eta ^{a}$. In (\ref{h2}) and further, $%
M_{ab}^{c}$ takes the form (\ref{c26}). Because all the components $%
\stackrel{(k)}{\omega }_{1}$ are known, it follows that $h_{1}$ is
completely determined.

Regarding the second-order deformation, we observe that the third term in
the equation (\ref{d10}) vanishes due to the fact that $\Omega _{2}=0$.
Making use of (\ref{h1}) and employing Jacobi's identity with respect to the
Dirac bracket, it is easy to see that the second term in (\ref{d10}) turns
into
\begin{equation}
\left[ H_{1},\Omega _{1}\right] =\frac{1}{2}\left[ K_{0},\left[ \Omega
_{1},\Omega _{1}\right] \right] ,  \label{h3}
\end{equation}
which vanishes due to the result established in the previous section,
according to which $\left[ \Omega _{1},\Omega _{1}\right] =0$. Then, we can
set
\begin{equation}
H_{2}=0,  \label{h5}
\end{equation}
which attracts that the remaining equations are satisfied for
\begin{equation}
H_{k}=0,\;k>2.  \label{h4}
\end{equation}
As a consequence, we can write the fully deformed BRST-invariant Hamiltonian
like
\begin{equation}
H_{\mathrm{B}}=H_{0\mathrm{B}}+gH_{1},  \label{h6}
\end{equation}
but also, taking into account (\ref{f31}), (\ref{c29}) and (\ref{h1})
\begin{equation}
H_{\mathrm{B}}=\left[ K_{0},\Omega \right] .  \label{h7}
\end{equation}
The last formula confirms the topological behaviour of the interacting
model. It stresses that $H_{\mathrm{B}}$ is not only invariant with respect
to the deformed Hamiltonian BRST symmetry, but also exact. This ends the
deformation of the BRST-invariant Hamiltonian for the free theory under
study.

\section{Identification of the interacting theory}

With the deformed BRST charge and BRST-invariant Hamiltonian at hand, in the
sequel we identify the Hamiltonian formulation of the interacting
first-class theory. Putting together the results from the previous two
sections, it follows that the complete expression of the deformed BRST
charge consistent to all orders in the deformation parameter is
\begin{eqnarray}
&&\Omega =\int d^{n-1}x\left( \eta ^{(1)a}\pi _{a}^{0}+2\eta _{a}^{(1)ij}\pi
_{ij}^{a}-C_{i}^{(1)a}p_{a}^{i}\right.  \nonumber \\
&&+C_{i}^{a}\left( \partial ^{i}\varphi _{a}+gW_{ab}A^{bi}\right)  \nonumber
\\
&&+\eta ^{a}\left( -\partial _{i}B_{a}^{0i}+gW_{ab}H_{0}^{b}-g\frac{\delta
W_{ab}}{\delta \varphi _{c}}A_{i}^{b}B_{c}^{0i}\right)  \nonumber \\
&&+\eta _{a}^{ij}\left( -\partial _{[i}A_{j]}^{a}-g\frac{\delta W_{bc}}{%
\delta \varphi _{a}}A_{i}^{b}A_{j}^{c}\right)  \nonumber \\
&&+C_{ij}^{a}\left( -\partial ^{[i}P_{a}^{j]}+g\frac{\delta W_{ab}}{\delta
\varphi _{c}}P_{c}^{[i}A^{bj]}-gW_{ab}\mathcal{P}^{bij}\right)  \nonumber \\
&&+\eta _{a}^{ijk}\left( \partial _{[i}\mathcal{P}_{jk]}^{a}-g\frac{\delta
W_{bc}}{\delta \varphi _{a}}\mathcal{P}_{[ij}^{b}A_{k]}^{c}+g\frac{\delta
^{2}W_{bc}}{\delta \varphi _{a}\delta \varphi _{d}}%
P_{d[i}A_{j}^{b}A_{k]}^{c}\right)  \nonumber \\
&&-g\frac{\delta W_{ab}}{\delta \varphi _{c}}P_{c}^{i}\eta ^{a}C_{i}^{b}-g%
\frac{\delta W_{ab}}{\delta \varphi _{c}}\left( \frac{1}{2}\eta ^{a}\eta ^{b}%
\mathcal{P}_{c}+\mathcal{P}_{ij}^{a}\eta ^{b}\eta _{c}^{ij}\right)  \nonumber
\\
&&+g\frac{\delta ^{2}W_{bc}}{\delta \varphi _{a}\delta \varphi _{d}}%
P_{di}\left( \frac{1}{2}\eta ^{b}\eta ^{c}B_{a}^{0i}+2A_{j}^{b}\eta ^{c}\eta
_{a}^{ij}\right)  \nonumber \\
&&+C_{ijk}^{a}\left( \partial ^{[i}P_{a}^{jk]}+g\frac{\delta W_{ab}}{\delta
\varphi _{c}}P_{c}^{[ij}A_{b}^{k]}+gW_{ab}\mathcal{P}^{bijk}\right.
\nonumber \\
&&\left. +g\frac{\delta W_{ab}}{\delta \varphi _{c}}\mathcal{P}%
^{b[ij}P_{c}^{k]}+g\frac{\delta ^{2}W_{ab}}{\delta \varphi _{c}\delta
\varphi _{d}}P_{c}^{[i}P_{d}^{j}A^{bk]}\right)  \nonumber \\
&&+\eta _{a}^{ijkl}\left( -\partial _{[i}\mathcal{P}_{jkl]}^{a}+g\frac{%
\delta W_{bc}}{\delta \varphi _{a}}A_{[i}^{c}\mathcal{P}_{jkl]}^{b}+g\frac{%
\delta W_{bc}}{\delta \varphi _{a}}\mathcal{P}_{[ij}^{b}\mathcal{P}%
_{kl]}^{c}\right.  \nonumber \\
&&+6g\frac{\delta ^{2}W_{bc}}{\delta \varphi _{a}\delta \varphi _{d}}P_{d[i}%
\mathcal{P}_{jk}^{b}A_{l]}^{c}-g\frac{\delta ^{2}W_{bc}}{\delta \varphi
_{a}\delta \varphi _{d}}P_{d[ij}A_{k}^{b}A_{l]}^{c}  \nonumber \\
&&\left. -g\frac{\delta ^{3}W_{bc}}{\delta \varphi _{a}\delta \varphi
_{d}\delta \varphi _{e}}P_{d[i}P_{ej}A_{k}^{b}A_{l]}^{c}\right)  \nonumber \\
&&-g\left( \frac{\delta W_{ab}}{\delta \varphi _{c}}P_{c}^{ij}+\frac{\delta
^{2}W_{ab}}{\delta \varphi _{c}\delta \varphi _{d}}P_{c}^{i}P_{d}^{j}\right)
\eta ^{a}C_{ij}^{b}  \nonumber \\
&&-g\left( \frac{\delta W_{ab}}{\delta \varphi _{c}}\mathcal{P}_{ijk}^{a}-%
\frac{\delta ^{2}W_{ab}}{\delta \varphi _{c}\delta \varphi _{d}}P_{d[i}%
\mathcal{P}_{jk]}^{a}\right) \eta ^{b}\eta _{c}^{ijk}  \nonumber \\
&&+g\left( \frac{\delta ^{2}W_{ab}}{\delta \varphi _{c}\delta \varphi _{d}}%
P_{dij}+\frac{\delta ^{3}W_{ab}}{\delta \varphi _{c}\delta \varphi
_{d}\delta \varphi _{e}}P_{di}P_{ej}\right) \times  \nonumber \\
&&\times \left( 3\eta ^{a}A_{k}^{b}\eta _{c}^{ijk}+\frac{1}{2}\eta ^{a}\eta
^{b}\eta _{c}^{ij}\right)  \nonumber \\
&&+\sum\limits_{k=5}^{n-1}\left( -\right) ^{k-1}\eta _{a}^{i_{1}i_{2}\cdots
i_{k}}\partial _{[i_{1}}\mathcal{P}_{i_{2}\cdots i_{k}]}^{a}  \nonumber \\
&&\left. +\sum\limits_{k=4}^{n-1}\left( -\right) ^{k-1}C_{i_{1}i_{2}\cdots
i_{k}}^{a}\partial ^{[i_{1}}P_{a}^{i_{2}\cdots
i_{k}]}+g\sum\limits_{k=3}^{n-1}\stackrel{(k)}{\omega }_{1}\right) ,
\label{i1}
\end{eqnarray}
while the full BRST-invariant Hamiltonian can be written in the form
\begin{eqnarray}
&&H_{\mathrm{B}}=\int d^{n-1}x\left( -H_{i}^{a}\left( \partial ^{i}\varphi
_{a}+gW_{ab}A^{bi}\right) \right.  \nonumber \\
&&+\frac{1}{2}B_{a}^{ij}\left( -\partial _{[i}A_{j]}^{a}-g\frac{\delta W_{bc}%
}{\delta \varphi _{a}}A_{i}^{b}A_{j}^{c}\right)  \nonumber \\
&&+A_{0}^{a}\left( -\partial _{i}B_{a}^{0i}+gW_{ab}H_{0}^{b}-g\frac{\delta
W_{ab}}{\delta \varphi _{c}}A_{i}^{b}B_{c}^{0i}\right)  \nonumber \\
&&+\eta ^{(1)a}\mathcal{P}_{a}+\eta _{a}^{(1)ij}\mathcal{P}%
_{ij}^{a}+C_{i}^{(1)a}P_{a}^{i}  \nonumber \\
&&+g\frac{\delta W_{ab}}{\delta \varphi _{c}}P_{c}^{i}\left( H_{i}^{a}\eta
^{b}+C_{i}^{a}A_{0}^{b}\right)  \nonumber \\
&&-g\frac{\delta W_{ab}}{\delta \varphi _{c}}\left( \frac{1}{2}%
B_{c}^{ij}\eta ^{a}\mathcal{P}_{ij}^{b}+A_{0}^{a}\mathcal{P}_{ij}^{b}\eta
_{c}^{ij}+A_{0}^{a}\eta ^{b}\mathcal{P}_{c}\right)  \nonumber \\
&&+g\frac{\delta ^{2}W_{ab}}{\delta \varphi _{c}\delta \varphi _{d}}%
P_{di}\left( B_{c}^{ij}A_{j}^{b}\eta ^{a}-\eta
^{a}A_{0}^{b}B_{c}^{0i}+2A_{0}^{a}A_{j}^{b}\eta _{c}^{ij}\right)  \nonumber
\\
&&+\frac{g}{4}\left( \frac{\delta ^{2}W_{ab}}{\delta \varphi _{c}\delta
\varphi _{d}}P_{d}^{ij}+\frac{\delta ^{3}W_{ab}}{\delta \varphi _{c}\delta
\varphi _{d}\delta \varphi _{e}}P_{d}^{i}P_{e}^{j}\right) \eta ^{a}\eta
^{b}B_{cij}  \nonumber \\
&&\left. +g\sum\limits_{k=2}^{n-1}A_{0}^{a}\frac{\partial ^{L}\stackrel{(k)}{%
\omega }_{1}}{\partial \eta ^{a}}\right) .  \label{i2}
\end{eqnarray}

By virtue of the discussion from Section 3 on the significance of the
various terms in the BRST charge and BRST-invariant Hamiltonian, at this
stage we extract the general features of the coupled model. Thus, the terms
of antighost number zero in (\ref{i1}) indicate that only the secondary
constraints are deformed as
\begin{equation}
\bar{G}_{a}^{(2)}\equiv -\left( D_{i}\right)
_{a}^{\;\;b}B_{b}^{0i}+gW_{ab}H_{0}^{b}\approx 0,  \label{i3}
\end{equation}
\begin{equation}
\bar{G}_{ij}^{(2)a}\equiv -\bar{F}_{ij}^{a}\approx 0,  \label{i4}
\end{equation}
\begin{equation}
\bar{\gamma}_{a}^{(2)i}\equiv D^{i}\varphi _{a}\approx 0,  \label{i5}
\end{equation}
where we employed the notations
\begin{equation}
\left( D_{i}\right) _{a}^{\;\;b}=\delta _{a}^{b}\partial _{i}+g\frac{\delta
W_{ac}}{\delta \varphi _{b}}A_{i}^{c},  \label{i6}
\end{equation}
\begin{equation}
\bar{F}_{ij}^{a}=\partial _{[i}A_{j]}^{a}+g\frac{\delta W_{bc}}{\delta
\varphi _{a}}A_{i}^{b}A_{j}^{c},  \label{i7}
\end{equation}
\begin{equation}
D^{i}\varphi _{a}=\partial ^{i}\varphi _{a}+gW_{ab}A^{bi}.  \label{i8}
\end{equation}
It is known that the first-class constraints generate gauge transformations.
In consequence, the gauge transformations of the interacting theory will
change with respect to the initial ones. From the pieces linear in the
antighost number one antighosts, as well as quadratic in the pure ghost
number one ghosts, we withdraw that some of the Dirac brackets among the new
constraint functions are modified as
\begin{equation}
\left[ \bar{G}_{a}^{(2)},\bar{G}_{b}^{(2)}\right] =-g\left( \frac{\delta
W_{ab}}{\delta \varphi _{c}}\bar{G}_{c}^{(2)}-\frac{\delta ^{2}W_{ab}}{%
\delta \varphi _{c}\delta \varphi _{d}}B_{d0i}\bar{\gamma}_{c}^{(2)i}\right)
,  \label{i9}
\end{equation}
\begin{equation}
\left[ \bar{G}_{a}^{(2)},\bar{G}_{ij}^{(2)b}\right] =g\left( \frac{\delta
W_{ac}}{\delta \varphi _{b}}\bar{G}_{ij}^{(2)c}-\frac{\delta ^{2}W_{ac}}{%
\delta \varphi _{b}\delta \varphi _{d}}\bar{\gamma}_{d[i}^{(2)}A_{j]}^{c}%
\right) ,  \label{i10}
\end{equation}
\begin{equation}
\left[ \bar{G}_{a}^{(2)},\bar{\gamma}_{b}^{(2)i}\right] =-g\frac{\delta
W_{ab}}{\delta \varphi _{c}}\bar{\gamma}_{c}^{(2)i},  \label{i11}
\end{equation}
so the gauge algebra of the first-class constraints is non-abelian and,
moreover, open. From the elements simultaneously linear in the pure ghost
number two ghosts and in the antighost number one antighosts we determine
the first-stage reducibility relations
\begin{equation}
\left( \bar{Z}_{i_{1}i_{2}i_{3}}^{a}\right) _{b}^{ij}\bar{G}%
_{ij}^{(2)b}+\left( \bar{Z}_{i_{1}i_{2}i_{3}}^{a}\right) _{i}^{b}\bar{\gamma}%
_{b}^{(2)i}=0,  \label{i12}
\end{equation}
\begin{equation}
\left( \bar{Z}_{a}^{i_{1}i_{2}}\right) _{b}^{ij}\bar{G}_{ij}^{(2)b}+\left(
\bar{Z}_{a}^{i_{1}i_{2}}\right) _{i}^{b}\bar{\gamma}_{b}^{(2)i}=0,
\label{i13}
\end{equation}
where the accompanying reducibility functions read as
\begin{equation}
\left( \bar{Z}_{i_{1}i_{2}i_{3}}^{a}\right) _{b}^{ij}=\frac{1}{2}\left(
D_{[i_{1}}\right) _{\;\;b}^{a}\delta _{i_{2}}^{i}\delta _{i_{3}]}^{j},
\label{i14}
\end{equation}
\begin{equation}
\left( \bar{Z}_{i_{1}i_{2}i_{3}}^{a}\right) _{i}^{b}=g\frac{\delta ^{2}W_{cd}%
}{\delta \varphi _{a}\delta \varphi _{b}}%
g_{i[i_{1}}A_{i_{2}}^{c}A_{i_{3}]}^{d},  \label{i14a}
\end{equation}
\begin{equation}
\left( \bar{Z}_{a}^{i_{1}i_{2}}\right) _{b}^{ij}=-\frac{1}{2}gW_{ab}\left(
g^{i_{1}i}g^{i_{2}j}-g^{i_{1}j}g^{i_{2}i}\right) ,  \label{i15a}
\end{equation}
\begin{equation}
\left( \bar{Z}_{a}^{i_{1}i_{2}}\right) _{i}^{b}=-\left( D^{[i_{1}}\right)
_{a}^{\;\;b}\delta _{i}^{i_{2}]},  \label{i15}
\end{equation}
with
\begin{equation}
\left( D_{i}\right) _{\;\;b}^{a}=\delta _{b}^{a}\partial _{i}-g\frac{\delta
W_{bc}}{\delta \varphi _{a}}A_{i}^{c}.  \label{i16}
\end{equation}
The part linear in the ghosts with pure ghost number $k+1\geq 3$ contains
polynomials of antighost number $k\geq 2$ more than linear in the
antighosts, which shows that the reducibility relations of order $k\geq 2$
hold on-shell. Indeed, from the inspection of this type of expressions, we
find at pure ghost number three ($k+1=3$) the on-shell second-stage
reducibility relations
\begin{eqnarray}
&&\left( \bar{Z}_{i_{1}i_{2}i_{3}i_{4}}^{a}\right)
_{b}^{j_{1}j_{2}j_{3}}\left( \bar{Z}_{j_{1}j_{2}j_{3}}^{b}\right)
_{c}^{ij}f_{ij}^{c}+\left( \bar{Z}_{i_{1}i_{2}i_{3}i_{4}}^{a}\right)
_{j_{1}j_{2}}^{b}\left( \bar{Z}_{b}^{j_{1}j_{2}}\right) _{c}^{ij}f_{ij}^{c}
\nonumber \\
&=&-g\left( \frac{\delta W_{bc}}{\delta \varphi _{a}}\bar{G}%
_{[i_{1}i_{2}}^{(2)b}f_{i_{3}i_{4}]}^{c}-\frac{\delta ^{2}W_{cd}}{\delta
\varphi _{a}\delta \varphi _{b}}\bar{\gamma}%
_{b[i_{1}}^{(2)}A_{i_{2}}^{c}f_{i_{3}i_{4}]}^{d}\right) ,  \label{i17}
\end{eqnarray}
\begin{eqnarray}
&&\left( \bar{Z}_{a}^{i_{1}i_{2}i_{3}}\right) _{j_{1}j_{2}}^{b}\left( \bar{Z}%
_{b}^{j_{1}j_{2}}\right) _{i}^{c}f_{c}^{i}+\left( \bar{Z}%
_{a}^{i_{1}i_{2}i_{3}}\right) _{b}^{j_{1}j_{2}j_{3}}\left( \bar{Z}%
_{j_{1}j_{2}j_{3}}^{b}\right) _{i}^{c}f_{c}^{i}  \nonumber \\
&=&g\left( \frac{\delta W_{ab}}{\delta \varphi _{c}}\bar{G}%
^{(2)b[i_{1}i_{2}}f_{c}^{i_{3}]}-\frac{\delta ^{2}W_{ab}}{\delta \varphi
_{c}\delta \varphi _{d}}\bar{\gamma}_{c}^{(2)[i_{1}}A^{bi_{2}}f_{d}^{i_{3}]}%
\right) ,  \label{i18}
\end{eqnarray}
where $f_{c}^{i}$ and $f_{ij}^{c}$ are arbitrary smooth functions (the
latter are antisymmetric in their spatial indices), along with the
second-stage reducibility functions
\begin{equation}
\left( \bar{Z}_{i_{1}i_{2}i_{3}i_{4}}^{a}\right) _{b}^{j_{1}j_{2}j_{3}}=-%
\frac{1}{3!}\left( D_{[i_{1}}\right) _{\;\;b}^{a}\delta
_{i_{2}}^{j_{1}}\delta _{i_{3}}^{j_{2}}\delta _{i_{4}]}^{j_{3}},  \label{i19}
\end{equation}
\begin{equation}
\left( \bar{Z}_{i_{1}i_{2}i_{3}i_{4}}^{a}\right) _{j_{1}j_{2}}^{b}=-\frac{g}{%
2}g_{j_{1}k_{1}}g_{j_{2}k_{2}}\frac{\delta ^{2}W_{cd}}{\delta \varphi
_{a}\delta \varphi _{b}}\delta _{[i_{1}}^{k_{1}}\delta
_{i_{2}}^{k_{2}}A_{i_{3}}^{c}A_{i_{4}]}^{d},  \label{i20}
\end{equation}
\begin{equation}
\left( \bar{Z}_{a}^{i_{1}i_{2}i_{3}}\right) _{j_{1}j_{2}}^{b}=\frac{1}{2}%
\left( D^{[i_{1}}\right) _{a}^{\;\;b}\delta _{j_{1}}^{i_{2}}\delta
_{j_{2}}^{i_{3}]},  \label{i21}
\end{equation}
\begin{equation}
\left( \bar{Z}_{a}^{i_{1}i_{2}i_{3}}\right) _{b}^{j_{1}j_{2}j_{3}}=\frac{g}{%
3!}W_{ab}\sum\limits_{\sigma \in S_{3}}\left( -\right) ^{\sigma
}g^{i_{1}j_{\sigma (1)}}g^{i_{2}j_{\sigma (2)}}g^{i_{3}j_{\sigma (3)}}.
\label{i22}
\end{equation}
In (\ref{i22}) $S_{3}$ signifies the set of permutations of $\{1,2,3\}$, and
$\left( -\right) ^{\sigma }$ means the parity of a certain permutation $%
\sigma $ pertaining to $S_{3}$. By making a similar analysis with respect to
the terms linear in the pure ghost number $\left( p+1\right) $ ghosts ($%
p=3,\cdots ,n-3$), we extract the on-shell $p$-stage reducibility relations
\begin{eqnarray}
&&\left( \bar{Z}_{i_{1}\cdots i_{p+2}}^{a}\right) _{b}^{j_{1}\cdots
j_{p+1}}\left( \bar{Z}_{j_{1}\cdots j_{p+1}}^{b}\right) _{c}^{k_{1}\cdots
k_{p}}  \nonumber \\
&&+\left( \bar{Z}_{i_{1}\cdots i_{p+2}}^{a}\right) _{j_{1}\cdots
j_{p}}^{b}\left( \bar{Z}_{b}^{j_{1}\cdots j_{p}}\right) _{c}^{k_{1}\cdots
k_{p}}\approx 0,  \label{i23}
\end{eqnarray}
\begin{eqnarray}
&&\left( \bar{Z}_{a}^{i_{1}\cdots i_{p+1}}\right) _{j_{1}\cdots
j_{p}}^{b}\left( \bar{Z}_{b}^{j_{1}\cdots j_{p}}\right) _{k_{1}\cdots
k_{p-1}}^{c}  \nonumber \\
&&+\left( \bar{Z}_{a}^{i_{1}\cdots i_{p+1}}\right) _{b}^{j_{1}\cdots
j_{p+1}}\left( \bar{Z}_{j_{1}\cdots j_{p+1}}^{b}\right) _{k_{1}\cdots
k_{p-1}}^{c}\approx 0,  \label{i24}
\end{eqnarray}
plus the $p$-stage reducibility functions
\begin{equation}
\left( \bar{Z}_{i_{1}\cdots i_{p+2}}^{a}\right) _{b}^{j_{1}\cdots j_{p+1}}=%
\frac{\left( -\right) ^{p+1}}{\left( p+1\right) !}\left( D_{[i_{1}}\right)
_{\;\;b}^{a}\delta _{i_{2}}^{j_{1}}\cdots \delta _{i_{p+2}]}^{j_{p+1}},
\label{i25}
\end{equation}
\begin{equation}
\left( \bar{Z}_{i_{1}\cdots i_{p+2}}^{a}\right) _{j_{1}\cdots j_{p}}^{b}=-%
\frac{\left( -\right) ^{p}g}{p!}g_{j_{1}k_{1}}\cdots g_{j_{p}k_{p}}\frac{%
\delta ^{2}W_{cd}}{\delta \varphi _{a}\delta \varphi _{b}}\delta
_{[i_{1}}^{k_{1}}\cdots \delta
_{i_{p}}^{k_{p}}A_{i_{p+1}}^{c}A_{i_{p+2}]}^{d},  \label{i26}
\end{equation}
\begin{equation}
\left( \bar{Z}_{a}^{i_{1}\cdots i_{p+1}}\right) _{j_{1}\cdots j_{p}}^{b}=%
\frac{\left( -\right) ^{p}}{p!}\left( D^{[i_{1}}\right) _{a}^{\;\;b}\delta
_{j_{1}}^{i_{2}}\cdots \delta _{j_{p}}^{i_{p+1}]},  \label{i27}
\end{equation}
\begin{eqnarray}
&&\left( \bar{Z}_{a}^{i_{1}\cdots i_{p+1}}\right) _{b}^{j_{1}\cdots j_{p+1}}=
\nonumber \\
&&\frac{\left( -\right) ^{p}g}{\left( p+1\right) !}W_{ab}\sum\limits_{\sigma
\in S_{p+1}}\left( -\right) ^{\sigma }g^{i_{1}j_{\sigma
(1)}}g^{i_{2}j_{\sigma (2)}}\cdots g^{i_{p+1}j_{\sigma (p+1)}}.  \label{i28}
\end{eqnarray}
In (\ref{i28}) $S_{p+1}$ and $\left( -\right) ^{\sigma }$ denote the set of
permutations of $\{1,2,\cdots ,p+1\}$, respectively, the parity of a
permutation $\sigma $ pertaining to $S_{p+1}$. Finally, the elements linear
in the pure ghost number $\left( n-1\right) $ ghosts describe the
reducibility relations of highest order
\begin{eqnarray}
&&\left( \bar{Z}_{a}^{i_{1}\cdots i_{n-1}}\right) _{j_{1}\cdots
j_{n-2}}^{b}\left( \bar{Z}_{b}^{j_{1}\cdots j_{n-2}}\right) _{k_{1}\cdots
k_{n-3}}^{c}f_{c}^{k_{1}\cdots k_{n-3}}  \nonumber \\
&&+\left( \bar{Z}_{a}^{i_{1}\cdots i_{n-1}}\right) _{b}^{j_{1}\cdots
j_{n-1}}\left( \bar{Z}_{j_{1}\cdots j_{n-1}}^{b}\right) _{k_{1}\cdots
k_{n-3}}^{c}f_{c}^{k_{1}\cdots k_{n-3}}=  \nonumber \\
&&-g\left( \frac{\delta ^{2}W_{ab}}{\delta \varphi _{c}\delta \varphi _{d}}%
\bar{\gamma}_{c}^{(2)[i_{1}}A^{bi_{2}}f_{d}^{i_{3}\cdots i_{n-1}]}-\frac{%
\delta W_{ab}}{\delta \varphi _{c}}\bar{G}^{(2)b[i_{1}i_{2}}f_{c}^{i_{3}%
\cdots i_{n-1}]}\right) ,  \label{i29}
\end{eqnarray}
\begin{eqnarray}
&&\left( \bar{Z}_{a}^{i_{1}\cdots i_{n-1}}\right) _{j_{1}\cdots
j_{n-2}}^{b}\left( \bar{Z}_{b}^{j_{1}\cdots j_{n-2}}\right)
_{c}^{k_{1}\cdots k_{n-2}}f_{k_{1}\cdots k_{n-2}}^{c}  \nonumber \\
&&+\left( \bar{Z}_{a}^{i_{1}\cdots i_{n-1}}\right) _{b}^{j_{1}\cdots
j_{n-1}}\left( \bar{Z}_{j_{1}\cdots j_{n-1}}^{b}\right) _{c}^{k_{1}\cdots
k_{n-2}}f_{k_{1}\cdots k_{n-2}}^{c}=  \nonumber \\
&&-g\frac{\delta W_{ab}}{\delta \varphi _{c}}\bar{\gamma}%
_{c}^{(2)[i_{1}}f^{bi_{2}\cdots i_{n-1}]},  \label{i30}
\end{eqnarray}
where $f_{c}^{k_{1}\cdots k_{n-3}}$ and $f_{k_{1}\cdots k_{n-2}}^{c}$ are
arbitrary completely antisymmetric smooth functions, and equally furnish the
$\left( n-2\right) $-order reducibility functions
\begin{equation}
\left( \bar{Z}_{a}^{i_{1}\cdots i_{n-1}}\right) _{j_{1}\cdots j_{n-2}}^{b}=%
\frac{\left( -\right) ^{n}}{\left( n-2\right) !}\left( D^{[i_{1}}\right)
_{a}^{\;\;b}\delta _{j_{1}}^{i_{2}}\cdots \delta _{j_{n-2}}^{i_{n-1}]},
\label{i31}
\end{equation}
\begin{eqnarray}
&&\left( \bar{Z}_{a}^{i_{1}\cdots i_{n-1}}\right) _{b}^{j_{1}\cdots j_{n-1}}=
\nonumber \\
&&\frac{\left( -\right) ^{n}g}{\left( n-1\right) !}W_{ab}\sum\limits_{\sigma
\in S_{n-1}}\left( -\right) ^{\sigma }g^{i_{1}j_{\sigma
(1)}}g^{i_{2}j_{\sigma (2)}}\cdots g^{i_{n-1}j_{\sigma (n-1)}}.  \label{i32}
\end{eqnarray}
The notations $S_{n-1}$ and $\left( -\right) ^{\sigma }$ are similar with
the above ones. This is of course not all the information we gain on the
interacting first-class theory. We actually know everything on the tensor
structure of the deformed first-class constraints from (\ref{i1}) if we
merely separate specific polynomials in the ghosts and antighosts. For
example, the relations (\ref{i9}--\ref{i11}) underline that the gauge
algebra of the deformed first-class constraints is open, and, meanwhile,
display the concrete form of the first-order structure functions. However,
there is a tower of higher-order structure functions, that satisfy recursive
equations, dictated in the Hamiltonian formulation by taking their repeated
Dirac brackets with the first-class constraint functions. These equations
will have an intricate form due to the fact that the interacting model is
also on-shell reducible. From (\ref{i1}) we can precisely withdraw these
higher-order structure functions, as well as the equations that relate them
at any level, if we just isolate the appropriate polynomials in the ghosts
and antighosts.

Now, we investigate the modified BRST-invariant Hamiltonian (\ref{i2}). The
component of antighost number zero
\begin{equation}
H=\int d^{n-1}x\left( -H_{i}^{a}\bar{\gamma}_{a}^{(2)i}+\frac{1}{2}B_{a}^{ij}%
\bar{G}_{ij}^{(2)a}+A_{0}^{a}\bar{G}_{a}^{(2)}\right) ,  \label{i33}
\end{equation}
represents nothing but the new first-class Hamiltonian, while the terms
linear in the antighost number one antighosts give the deformed gauge
algebra relations
\begin{equation}
\left[ H,G_{a}^{(1)}\right] =\bar{G}_{a}^{(2)},  \label{i34}
\end{equation}
\begin{eqnarray}
&&\left[ H,\bar{G}_{a}^{(2)}\right] =g\frac{\delta W_{ab}}{\delta \varphi
_{c}}\left( A_{0}^{b}\bar{G}_{c}^{(2)}-H_{i}^{b}\bar{\gamma}_{c}^{(2)i}-%
\frac{1}{2}\bar{G}_{ij}^{(2)b}B_{c}^{ij}\right)  \nonumber \\
&&+g\frac{\delta ^{2}W_{ac}}{\delta \varphi _{b}\delta \varphi _{d}}\left(
\frac{1}{2}B_{b}^{ij}\bar{\gamma}_{d[i}^{(2)}A_{j]}^{c}-B_{d0i}A_{0}^{c}\bar{%
\gamma}_{b}^{(2)i}\right) ,  \label{i35}
\end{eqnarray}
\begin{equation}
\left[ H,G_{ij}^{(1)a}\right] =\bar{G}_{ij}^{(2)a},  \label{i36}
\end{equation}
\begin{equation}
\left[ H,\bar{G}_{ij}^{(2)a}\right] =g\left( \frac{\delta W_{bc}}{\delta
\varphi _{a}}A_{0}^{b}\bar{G}_{ij}^{(2)c}-\frac{\delta ^{2}W_{cd}}{\delta
\varphi _{a}\delta \varphi _{b}}A_{0}^{c}\bar{\gamma}_{b[i}^{(2)}A_{j]}^{d}%
\right) ,  \label{i37}
\end{equation}
\begin{equation}
\left[ H,\gamma _{a}^{(1)i}\right] =\bar{\gamma}_{a}^{(2)i},  \label{i38}
\end{equation}
\begin{equation}
\left[ H,\bar{\gamma}_{a}^{(2)i}\right] =g\frac{\delta W_{ab}}{\delta
\varphi _{c}}A_{0}^{b}\bar{\gamma}_{c}^{(2)i}.  \label{i39}
\end{equation}
Just like in the case of the BRST charge, the formula (\ref{i2}) tells us
everything on the tensor structure of the interacting Hamiltonian (\ref{i33}%
) that governs the dynamics on the deformed first-class surface. Indeed,
from (\ref{i34}--\ref{i39}) we learn, besides the first-class behaviour of $%
H $, that there appear some nontrivial structure functions. This means that
there will also be a consequent recursive `open setting', formulated in
terms of higher-order structure functions of $H$ and of the equations that
relate them, that can be derived by taking the repeated Dirac brackets of (%
\ref{i34}--\ref{i39}) with the deformed first-class constraint functions. Of
course, the on-shell reducibility of the new first-class constraints will be
involved at every stage. We have this entire setting at our hand, and can
write it down at any level, simply by identifying the adequate polynomials
in the ghosts and antighosts from (\ref{i2}).

So far, it is clear that the entire Hamiltonian deformation is controlled by
$W_{ab}$ since if we set $W_{ab}=0$ we recover the initial free topological
field theory even when the coupling constant is different from zero. (In
other words, we get no deformations at all.) Moreover, we have seen that the
consistency of the deformation restricts $W_{ab}\left( \varphi \right) $ to
be antisymmetric and to satisfy the identity (\ref{c27}). Let us see the
geometric meaning of this so-called `two-tensor'. To this end, we briefly
review the basic notions on Poisson manifolds. If $N$ denotes an arbitrary
Poisson manifold, then this is equipped with a Poisson bracket $\left\{
,\right\} $ that is bilinear, antisymmetric, subject to a Leibnitz-like rule
and satisfies a Jacobi-type identity. If $\left\{ X^{i}\right\} $ are some
local coordinates on $N$, then there exists a two-tensor $\mathcal{P}%
^{ij}\equiv \left\{ X^{i},X^{j}\right\} $ (the Poisson tensor) that uniquely
determines the Poisson structure together with the Leibnitz rule. This
two-tensor is antisymmetric and transforms covariantly under coordinate
transformations. Jacobi's identity for the Poisson bracket $\left\{
,\right\} $ expressed in terms of the Poisson tensor reads as $\mathcal{P}%
_{,k}^{ij}\mathcal{P}^{kl}+\mathrm{cyclic}\left( i,j,l\right) =0$, where $%
\mathcal{P}_{,k}^{ij}\equiv \partial \mathcal{P}^{ij}/\partial X^{k}$. Now,
the geometric origin of $W_{ab}$ is obvious. If, for instance, we choose a
concrete form for the antisymmetric functions $W_{ab}\left( \varphi \right) $
that satisfy (\ref{c27}), then we can interpret the dynamical scalar fields $%
\left\{ \varphi _{a}\right\} $ precisely like some local coordinates on a
target manifold endowed with a prescribed Poisson structure (up to the plain
convention that the lower index $a$ is a `covariant' index of the type $i$).
Conversely, any given Poisson manifold parametrized in terms of some local
coordinates $\left\{ \varphi _{a}\right\} $ (within the same index
convention) prescribes a Poisson tensor $W_{ab}\left( \varphi \right) $
which is antisymmetric and satisfies (\ref{c27}). This discussion also
argues that the attribute of `two-tensor' given to $W_{ab}$ is not
misleading, but only hidden behind some Poisson structure. It is clear that
for an odd number of scalar fields, $W_{ab}$ is degenerate irrespective of
its concrete form. For an even number of scalar fields, we can find
nondegenerate forms of $W_{ab}$, whose inverse will be nothing but the
symplectic two-form on the target space, which becomes a symplectic manifold.

Passing to the Lagrangian formulation of the interacting theory, after some
computation we get the action
\begin{equation}
S\left[ A_{\mu }^{a},H_{\mu }^{a},\varphi _{a},B_{a}^{\mu \nu }\right] =\int
d^{n}x\left( H_{\mu }^{a}D^{\mu }\varphi _{a}+\frac{1}{2}B_{a}^{\mu \nu }%
\bar{F}_{\mu \nu }^{a}\right) ,  \label{i40}
\end{equation}
subject to the gauge invariances
\begin{equation}
\delta _{\epsilon }A_{\mu }^{a}=\left( D^{\mu }\right) _{\;\;b}^{a}\epsilon
^{b},  \label{i41}
\end{equation}
\begin{equation}
\delta _{\epsilon }\varphi _{a}=-gW_{ab}\epsilon ^{b},  \label{i42}
\end{equation}
\begin{eqnarray}
&&\delta _{\epsilon }H_{\mu }^{a}=\left( D^{\nu }\right)
_{\;\;b}^{a}\epsilon _{\mu \nu }^{b}-g\frac{\delta W_{bc}}{\delta \varphi
_{a}}\epsilon ^{b}H_{\mu }^{c}  \nonumber \\
&&+g\frac{\delta ^{2}W_{cd}}{\delta \varphi _{a}\delta \varphi _{b}}\left(
\frac{1}{2}A^{c\nu }A^{d\rho }\epsilon _{b\mu \nu \rho }+A^{d\nu }\epsilon
^{c}B_{b\mu \nu }\right) ,  \label{i43}
\end{eqnarray}
\begin{equation}
\delta _{\epsilon }B_{a}^{\mu \nu }=\left( D_{\rho }\right)
_{a}^{\;\;b}\epsilon _{b}^{\mu \nu \rho }+gW_{ab}\epsilon ^{b\mu \nu }-g%
\frac{\delta W_{ab}}{\delta \varphi _{c}}\epsilon ^{b}B_{c}^{\mu \nu },
\label{i44}
\end{equation}
where $D^{\mu }\varphi _{a}$, $\bar{F}_{\mu \nu }^{a}$, $\left( D^{\mu
}\right) _{\;\;b}^{a}$ and $\left( D_{\rho }\right) _{a}^{\;\;b}$ can be
read from the formulas (\ref{i6}--\ref{i8}) and (\ref{i16}) by manifest
Lorentz covariance. The deformation of the Lagrangian gauge transformations
roots in the deformed first-class constraints (\ref{i3}--\ref{i5}). It can
be shown that these gauge transformations are on-shell $\left( n-2\right) $%
-order reducible and give rise to an open gauge algebra.

We notice that neither the Lagrangian action, nor the gauge
symmetry of the interacting theory, do contain the $n$-dimensional
antisymmetric symbol. This is a direct consequence of the fact
that we have removed the term containing the spatial part of this
symbol from the last component in the first-order deformation of
the BRST charge. Indeed, such a term would have resulted at the
level of the Lagrangian action in the vertex
\begin{equation}
\varepsilon ^{\mu _{1}\cdots \mu _{n}}U_{a_1\cdots a_n}A^{a_1}
_{\mu _{1}}\cdots A^{a_n} _{\mu _{n}}, \label{i70}
\end{equation}
that breaks the PT-invariance. In (\ref{i70}), the functions
$U_{a_1\cdots a_n}$ involve only the undifferentiated scalar
fields, are completely antisymmetric in their indices, and are
required to satisfy some identities implied by the consistency of
the first-order deformation of the BRST charge. This type of
interactions will be reported elsewhere.

\section{Conclusion}

In conclusion, in this paper we have generated the consistent
Hamiltonian interactions in any spacetime dimension $n\geq 4$ that
can be introduced among a set of scalar fields, two types of
one-forms and a system of two-forms, pictured in the free limit by
an abelian topological field theory of BF-type. Our treatment is
mainly based on the Hamiltonian BRST deformation procedure, that
relies on the construction of the consistent deformations of both
BRST charge and BRST-invariant Hamiltonian of the free model with
the help of some cohomological techniques. In addition, we require
that the deformations are local and independent of the spacetime
dimension. The results regarding the deformation of the BRST
charge can be synthesized by the fact that only the first-order
deformation can be taken to be nonvanishing, while its consistency
reveals some functions on the undifferentiated scalar fields that
can be seen as the components of a Poisson two-tensor on the
target space. Concerning the deformation of the BRST-invariant
Hamiltonian, it stops at order one in the coupling constant as
well, and, moreover, is exact with respect to the deformed
Hamiltonian BRST symmetry (see (\ref{h7})). From these two
deformed quantities we derive the Hamiltonian formulation of the
resulting coupled model, namely, its first-class constraints,
accompanying reducibility functions, first-class Hamiltonian and
gauge algebra relations. This is an example of deformation that
modifies the gauge transformations, the reducibility relations,
and also the gauge algebra. The resulting model is included
precisely within the class of interacting topological field
theories of BF-type with an open Hamiltonian gauge algebra and
on-shell reducibility relations. This work generalizes our
previous results from \cite{mpla}--\cite{ijmpa} in the sense that,
although the gauge structure of the interacting model is richer,
the Lagrangian of the interacting theory has a similar expression.
We mention that the two-dimensional case studied in \cite{mpla} is
irreducible and, in fact, equivalent to the standard Poisson Sigma
Model \cite{stroblspec}, up to the fact that it is written in more
complicated variables, but this equivalence is no longer valid in
$n\geq 4$ dimensions, where it can be observed a complex structure
of new nontrivial terms.

\acknowledgments{This work has been financed from the type A
grant, code 943/2002, with the Romanian Council for Academic
Scientific Research (CNCSIS) and the Romanian Ministry of
Education and Research (MEC). The authors wish to thank Dr. Thomas
Strobl for his kind suggestions and comments, that helped at the
completion of this work.}

\end{document}